\documentclass[iop,apj,tighten,numberedappendix]{emulateapj}
\usepackage[breaklinks,colorlinks,urlcolor=blue,citecolor=blue,linkcolor=blue]{hyperref}
\usepackage{amsmath,amssymb}
\usepackage{cleveref}
\usepackage{graphicx}
\usepackage{bm}
\usepackage[toc,title,page]{appendix}
\usepackage{tocbibind}
\usepackage{color}
\newcommand{\Msun}{{\rm M}_\odot}

\shorttitle{Regulation of gas accretion}
\shortauthors{et al.}

\begin{document}
\title{
Can stellar-mass black hole growth disrupt disks of active galactic nuclei? 
\\The role of mechanical feedback
}
\author{
Hiromichi Tagawa\altaffilmark{1}, 
Shigeo S Kimura\altaffilmark{1},
Zolt\'an Haiman\altaffilmark{2},
Rosalba Perna\altaffilmark{3,4},
Hidekazu Tanaka\altaffilmark{1},
Imre Bartos\altaffilmark{5}, 
}
\affil{
\altaffilmark{1}Astronomical Institute, Graduate School of Science, Tohoku University, Aoba, Sendai 980-8578, Japan\\
\altaffilmark{2}Department of Astronomy, Columbia University, 550 W. 120th St., New York, NY, 10027, USA\\
\altaffilmark{3}Department of Physics and Astronomy, Stony Brook University, Stony Brook, NY 11794-3800, USA\\
\altaffilmark{4}Center for Computational Astrophysics, Flatiron Institute, New York, NY 10010, USA\\
\altaffilmark{5}{Department of Physics, University of Florida, PO Box 118440, Gainesville, FL 32611, USA}\\
}
\email{E-mail: htagawa@astr.tohoku.ac.jp}

\begin{abstract} 
Stellar-mass BHs (sBHs) are predicted to be embedded in active galactic nuclei (AGN) disks due to gravitational drag 
and in-situ star formation. 
However, we find that due to a high gas density 
in an AGN disk environment, 
compact objects may rapidly grow to intermediate-mass BHs and deplete matter from the AGN disk unless accretion is suppressed by some feedback process(es). 
These consequences are inconsistent with AGN observations and the dynamics of the Galactic center. 
Here we consider mechanical feedback mechanisms 
for the reduction of gas accretion. 
Rapidly accreting sBHs launch winds and/or jets via the Blandford-Znajek mechanism, which produce high-pressure shocks and cocoons. 
Such a shock and cocoon can spread laterally in the plane of the disk, eject the outer regions of a circum-sBH disk (CsBD) and puncture a hole in the AGN disk with horizontal size comparable to the disk scale-height. 
Since the depletion timescale of the bound CsBD is much shorter than the resupply timescale of gas to the sBH, the time-averaged accretion rate onto sBHs is reduced by this process by a factor of $\sim 10$--$100$. 
This feedback mechanism can therefore help alleviate the sBH over-growth and AGN-disk depletion problems. 
On the other hand, 
we find that cocoons of jets 
can unbind a large fraction of the gas accreting in the disks of less massive SMBHs, which may help explain the dearth of high-Eddington ratio AGNs with SMBH mass $\lesssim10^5{\rm M_\odot}$.
\end{abstract}

\section{Introduction}

Massive galaxies are observed to have supermassive black holes (SMBHs) in their centers \citep[e.g.][for a review]{Kormendy2013}. 
It is well established that 
SMBHs have mainly grown via gas accretion from the disks of active galactic nuclei (AGNs) at least at redshifts $z\lesssim 5$ \citep[e.g.][]{Yu2002}. 
Several studies have suggested that densely populated stars and compact objects, including stellar-mass BHs (sBHs) in nuclear star clusters 
\citep[e.g.][]{MiraldaEscude00,Lu13} are captured and embedded in AGN disks \citep{Ostriker1983,Syer1991}, and stars actively form in the outer regions of AGN disks
\citep{Goodman03,Thompson05,Nayakshin07}. 
There are several observations supporting this picture \citep{Artymowicz1993,Levin2003,Tagawa19}. 
Additionally, AGN disks are promising environments to explain the characteristic properties of some unexpected gravitational wave events discovered by LIGO/Virgo, such as the massive-binary merger event, GW190521 \citep[e.g.][]{LIGO20_GW190521,Samsing20,Tagawa20_ecc,Tagawa20_MassGap} and its hypothesized electromagnetic counterpart \citep{Graham20}. 
However, 
if sBHs in AGN disks accrete without any feedback 
\citep[e.g.][]{Lubow1999,Levin2007,Tanigawa2016,Dittmann2021}, 
sBHs rapidly grow to intermediate-mass BHs (IMBHs) or SMBHs within the lifetime of AGNs \citep{Goodman04} and deplete most accreting gas, which contradicts quasar observations \citep[e.g.][]{Yu2002} 
and the dynamics of the Galactic center \citep{Naoz2020,Gravity2020}.

There are several feedback processes which may suppress rapid accretion. 
One is mechanical feedback by 
winds launched from the inner region of a rapidly accreting circum-sBH disk (CsBD, 
e.g. \citealt{Jiang+2014,Sadowski+2015,Regan19}). 
However, the wind 
expected under conditions of super-Eddington accretion is predicted to be launched anisotropically \citep[e.g.][]{Jiao2015,Kitaki2021}, and 
the gas inflow outside the wind opening angle is only modestly suppressed 
as long as the wind is not decelerated in a shock \citep{Takeo2020}. 
If most of the inflowing gas is ejected as the wind 
\citep[e.g.][]{Sadowski+2015}, 
the feedback would be stronger, and 
the growth of sBHs 
can be suppressed. 
On the other hand, the outflow rate for super-Eddington accretion is suggested to be modest when the trapping radius is much smaller than the circularization radius \citep{Kitaki2021}, which is the case for accretion onto sBHs in AGN disks \citep[e.g.][]{Tanigawa2012}. 
Most of previous studies have not investigated the wind feedback for the situations in which the wind is thermalized. 
Although \citet{Kimura2021_BubblesBHMs} and 
\citet{Wang2021_TZW,Wang2021b} examined the evolution of bubbles around sBHs due to the winds, 
they did not estimate the amount of the CsBD and surrounding gas ejected by the wind bubbles and its global importance for the AGN.

Another is radiation pressure driven ejection \citep[e.g.][]{Inayoshi2016_HyperEdd,Toyouchi19}.
An optically and geometrically thick inner 
CsBD 
is formed for a super-Eddington accretion flow, the radiation from the inner regions escapes perpendicular to the disk, and gas inflow along the disk plane is not significantly suppressed \citep{Sugimura17,Toyouchi2021}. 
Additionally, in dense environments like AGN disks, rapid accretion is 
not suppressed by radiation even if it is isotropic \citep{Inayoshi2016_HyperEdd}.  
Thus, to solve the over-growth and depletion 
problems, additional feedback processes are likely required.

In this paper we 
focus on a 
regulation process for accretion, mediated by the evolution of a cocoon generated around a jet launched by an accreting and rotating sBH due to the Blandford-Znajek (BZ) mechanism \citep{Blandford1977}. 
In this process, the cocoon can interact with and eject 
the CsBD 
as it spreads laterally towards the disk plane from the jet head, due to its high pressure coupled with 
local density gradients in the AGN disk near the sBH. 
Such feedback process is often called a jet feedback mechanism (JFM), 
and plays important roles in several contexts \citep[e.g.][for a review]{Soker2016} 
such as 
galaxy clusters \citep[e.g.][]{McNamara2012}, 
galaxy formation \citep[e.g.][]{Fabian2012}, 
planetary nebulae \citep[e.g.][]{Balick2002}, 
common envelope evolution \citep[e.g.][]{Soker2014}, and young stellar objects \citep[e.g.][]{Frank2014}. 
Here, we apply the JFM to the system of accreting sBHs in AGN disks. 
The JFM in the AGN disk is similar to that in common envelope evolution \citep[e.g.][]{MorenoMendez2017,LopezCamara2019,Grichener2021}, while there are many differences (the density and geometry of surrounding gas, the angular momentum of the CsBD, and the jet luminosity). Especially we found that the JFM for accretion in the AGN disk can eject the outer regions of the CsBD, which has not appeared in the other contexts. 
Similarly, a wind launched from the CsBD 
can eject the outer CsBD through strong shocks.
However, we find that the CsBD is typically more  efficiently ejected by the 
JFM, 
due to its higher pressure compared to the wind shock. We therefore focus on the 
JFM 
in the main text, and comment on the relative importance of wind feedback in Appendix~$\ref{sec:wind_feedback}$. 
We find that this regulation process can reduce the accretion rate by a factor of $\sim 10$--$100$, depending on the model parameters. 
We use the cylindrical coordinates $(z,r)$ and $(Z,R)$, with $z=r=0$ and $Z=R=0$ representing the positions of the sBH and the SMBH, respectively. 
The $Z=0$ plane is set to the AGN plane, and  
the $z$ axis represents the direction of the jet propagation. 
Unless stated otherwise, we assume below that the $z$ axis is aligned with the $Z$ axis.

\begin{figure}
\begin{center}
\includegraphics[width=80mm]{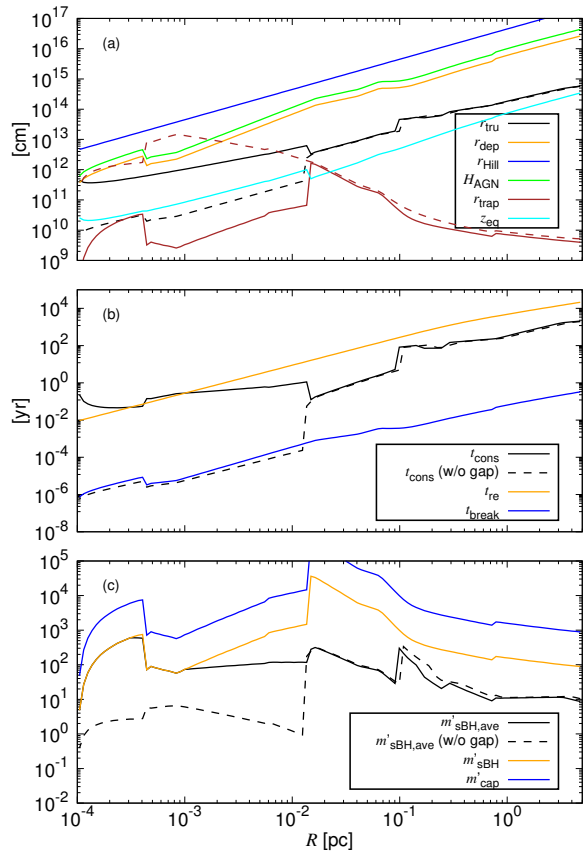}
\caption{
Several model quantities as a function of $R$ 
for the fiducial model ($\S\,\ref{sec:numerical_choice}$ and Table~\ref{table:parameter_fiducial}). 
{\em Upper} panel
(a): The truncation radius (black), depletion radius 
(orange), Hill radius (blue), AGN scale height (green), 
photon trapping radius (brown), 
and wind equilibrium height (cyan). {\em Middle} panel
(b): 
The timescales for CsBD gas consumption (black), resupply (orange), and jet breakout (blue). 
{\em Bottom} panel
(c):
The average accretion rate with the 
JFM 
(black), the
accretion rate (orange), 
and the
capture rate (blue) 
in units of the Eddington rate. 
In the case with dashed lines, a gap is assumed not to form in an AGN disk.} 
\label{fig:r_vari3}
\end{center}
\end{figure}

\begin{figure}
\begin{center}
\includegraphics[width=80mm]{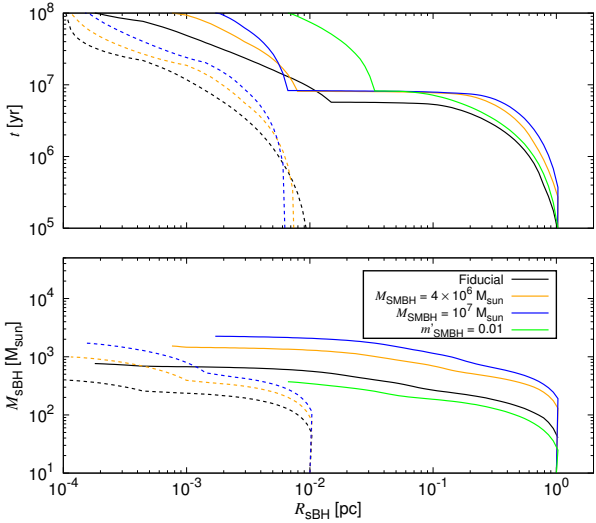}
\caption{
The time from the beginning of the AGN phase (upper panel) for the sBH to grow to the mass $M_{\rm sBH}$
(lower panel) 
for the fiducial model (black) as a function of the position $R_{\rm sBH}$, assumed to be initially at $R_{\rm ini}=1$~pc (solid lines) and 0.01 pc (dotted lines), respectively. 
The results for an SMBH mass of $M_{\rm SMBH}=4\times 10^6\,\Msun$ (orange), $10^7\,\Msun$ (blue), and for the Eddington accretion rate of ${\dot m}_{\rm SMBH}=0.01$ (green) are also shown. 
The parameters of the fiducial model are given in $\S\,\ref{sec:numerical_choice}$ and Table~\ref{table:parameter_fiducial}. 
}
\label{fig:r_m_noreg}
\end{center}
\end{figure}

\section{Rapid accretion problems}

\label{sec:rapid_growth}

We first outline the key issue related to the evolution of compact objects in AGN disks, that is a rapid growth of compact objects (e.g. \citealt{Goodman04}, see also \citealt{Cantiello2021,Jermyn2021} for stars). 
We estimate the growth rate of sBHs in an AGN disk without feedback as follows. 
For an sBH embedded in an AGN disk, the Bondi-Hoyle-Lyttleton radius ($r_{\rm BHL}$) usually exceeds the scale height of the AGN disk ($H_{\rm AGN}$) and the Hill radius ($r_{\rm Hill}$). Due to the limitation of capture regions by the shear motion and the vertical height of the AGN disk, 
the capture rate of gas by the sBH is given by
\begin{align}
\label{eq:md_bondi}
{\dot M}_{\rm cap}
=& f_{\rm c} r_{\rm w} r_{\rm h} \rho_{\rm AGN} (c_{\rm s,AGN}^2+v_{\rm sBH}^2+v_{\rm sh}^2)^{1/2}\nonumber\\
\simeq & ~3\times 10^{-4} \,\Msun/{\rm yr}~
\left(\frac{f_{\rm c}}{10}\right)
\left(\frac{H_{\rm AGN}}{0.003\,{\rm pc}}\right)
\left(\frac{R_{\rm sBH}}{1~{\rm pc}}\right)^{1/2}\nonumber\\
&
\left(\frac{\rho_{\rm AGN}}{4\times 10^{-17}\,{\rm g/cm^3}}\right)
\left(\frac{M_{\rm sBH}}{10~\Msun}\right)^{2/3}
\left(\frac{M_{\rm SMBH}}{10^6\,\Msun}\right)^{-1/6},
\end{align}
\citep[e.g.][]{Tanigawa2016,Stone17,Rosenthal2020}, 
where 
$\rho_{\rm AGN}$ is the gas density 
and $c_{\rm s,AGN}$ is the sound speed 
of the AGN disk at the position of the sBH ($R_{\rm sBH}$), 
$v_{\rm sBH}$ is the velocity of the sBH with respect to the local motion of the AGN disk, 
$v_{\rm sh}=r_{\rm w}(GM_{\rm SMBH}/R_{\rm sBH}^3)^{1/2}$ is the shear velocity 
at the capture radius 
$r_{\rm w}={\rm min}(r_{\rm BHL},\,r_{\rm Hill})$, 
$r_{\rm h}={\rm min}(r_{\rm w},\,H_{\rm AGN})$ is the capture height, 
$G$ is the gravitational constant, 
$M_{\rm SMBH}$ is the mass of the SMBH at the center of the AGN disk, 
and $f_{\rm c}$ is a normalization constant. 
We adopt $f_{\rm c}=10$ as found by \citet{Tanigawa2002}. 
In the second equality of Eq.~\eqref{eq:md_bondi}, we assume $v_{\rm sBH}<c_{\rm s,AGN}<v_{\rm sh}$ and 
$H_{\rm AGN}< r_{\rm Hill}< r_{\rm BHL}$.
Fig.~\ref{fig:r_vari3} shows various variables as a function of $R$, and as can be seen in panel~(a) of this figure, this assumption is satisfied in the disk model adopted in this paper. 
In the right side of Eq.~\eqref{eq:md_bondi} (and for the equations and figures below), the fiducial values for the model parameters ($\S\,\ref{sec:numerical_choice}$, Table~\ref{table:parameter_fiducial}) 
are adopted. The density and scale height of the AGN disk are derived from the model in \citet{Thompson05} as constructed in \citet{Tagawa19}. 
The accretion rate in units of the Eddington rate (${\dot M}_{\rm Edd}(M)$ for the mass $M$) with the conversion efficiency to radiation of $\eta_{\rm rad}=0.1$ is 
\begin{align}
\label{eq:md_bhl_edd}
{\dot m}_{\rm cap}= &
{\dot M}_{\rm cap}/
{\dot M}_{\rm Edd}(M_{\rm sBH})\nonumber\\
\simeq & 2\times 10^3
\left(\frac{f_{\rm c}}{10}\right)
\left(\frac{\eta_{\rm rad}}{0.1}\right)
\left(\frac{R_{\rm sBH}}{1\,{\rm pc}}\right)^{1/2}\nonumber\\
&\left(\frac{H_{\rm AGN}}{0.003~{\rm pc}}\right)
\left(\frac{M_{\rm sBH}}{10\,\Msun}\right)^{-1/3}
\left(\frac{M_{\rm SMBH}}{10^6\,\Msun}\right)^{-1/6}\nonumber\\
&\left(\frac{\rho_{\rm AGN}}{4\times 10^{-17}\,{\rm g/cm^3}}\right)
\end{align}
(blue line in Fig.~\ref{fig:r_vari3}~c). 
Also, sBHs are assumed to radially migrate in an AGN disk following the formulae developed for migration of planets in a proto-planetary disk (see Appendix~\ref{sec:migration}). 

Fig.~\ref{fig:r_m_noreg} shows 
the evolution of the sBH mass ($M_{\rm sBH}$), as well as the time ($t$) from the beginning of the AGN phase for the sBH to grow to $M_{\rm sBH}$, as a function of $R_{\rm sBH}$ for several combinations of values for $M_{\rm SMBH}$, $R_{\rm ini}$, and ${\dot m}_{\rm SMBH}$ (see Appendix~\ref{sec:bh_growth} for equations), 
where ${\dot m}_{\rm SMBH}={\dot M}_{\rm SMBH}/{\dot M}_{\rm Edd}(M_{\rm SMBH})$ is the gas inflow rate to the AGN disk in units of the Eddington rate for $M_{\rm SMBH}$.

Based on the sBH growth rate, we highlight two problems for the evolution of compact objects in AGN disks.

\begin{figure}
\begin{center}
\includegraphics[width=80mm]{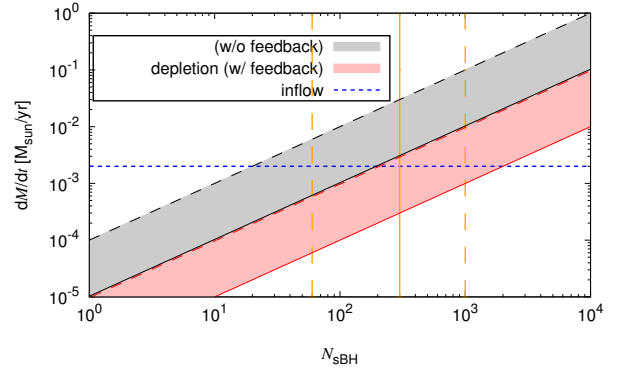}
\caption{
The depletion rate (${\dot M}$) with (red) and without (black) 
the JFM 
for $t_{\rm AGN}=10$ (dashed)--$100~{\rm Myr}$ (solid), 
and 
the inflow rate from the outer boundary of the AGN disk 
(blue dashed) 
as a function of the number of sBHs ($N_{\rm sBH}$) embedded in an AGN disk 
for $M_{\rm SMBH}=10^6\,\Msun$ and ${\dot m}_{\rm SMBH}=0.1$. 
The $y$- and $x$-axis scale as $(M_{\rm SMBH}/10^6\,\Msun)({\dot m}_{\rm SMBH}/0.1)$ and $({\dot m}_{\rm SMBH}/0.1)^{1/2}$, respectively. 
The orange vertical lines show 
the fiducial value (solid) and rough upper and lower limits (dashed) for $N_{\rm sBH}{\dot m}_{\rm SMBH,-1}^{1/2}$.  As the figure shows, the expected number, $60$--$1000$, of sBHs embedded in the AGN disk can cumulatively accrete a comparable amount, or even more, than the proposed SMBH accretion rate, thus starving the SMBH.
}
\label{fig:dep_m}
\end{center}
\end{figure}

\noindent{(i) "Depletion Problem".}

\noindent 
The number of captured sBHs ($N_{\rm sBH,AGN}$) can be estimated to be 
$N_{\rm AGN,sBH}\sim 60$--$1000\,({\dot m}_{\rm Edd}/0.1)^{1/2}$ 
\citep{Tagawa19,Tagawa20_MassGap}. This value is  
roughly consistent with the number of low-mass X-ray binaries 
 in the Galactic center region \citep{Hailey18,Tagawa19,Mori2021} and the rates for some fraction of sBH mergers detected by LIGO-Virgo (e.g. GW190521) originating in AGNs \citep{Tagawa20_MassGap}. 
By roughly assuming 
\begin{align}
\label{eq:n_agn_sbh}
N_{\rm AGN,sBH}\sim 300\,
\left(\frac{{\dot m}_{\rm SMBH}}{0.1}\right)^{1/2} 
\end{align}
and 
the mass of an sBH at the end of the AGN phase  
($t_{\rm AGN}\sim 10$--$100\,{\rm Myr}$, e.g., \citealt{Marconi04,Greene07}) 
to be 
\begin{align}
M_{\rm sBH,fin}\sim 10^3\,\Msun~
\left(\frac{M_{\rm SMBH}}{10^6\,\Msun}\right)
\left(\frac{{\dot m}_{\rm SMBH}}{0.1}\right)^{1/2}
\end{align}
(from Fig.~\ref{fig:r_m_noreg}), the gas in the AGN disk is depleted  by the population of sBHs embedded in it at 
a rate
\begin{align}
&{\dot M}_{\rm sBH,tot} \sim N_{\rm AGN,sBH}M_{\rm sBH,fin}/t_{\rm AGN}\nonumber\\
&\sim 3\times10^{-2}-3\times 10^{-3}\,\Msun/{\rm yr}~
\left(\frac{M_{\rm SMBH}}{10^6\,\Msun}\right)
\left(\frac{{\dot m}_{\rm SMBH}}{0.1}\right).
\end{align}
This exceeds the inflow rate of the AGN disk, 
\begin{align}
{\dot M}_{\rm SMBH,in}\simeq 2\times 10^{-3}\,\Msun/{\rm yr}~
\left(\frac{M_{\rm SMBH}}{10^6\,\Msun}\right)
\left(\frac{{\dot m}_{\rm SMBH}}{0.1}\right),
\end{align}
implying that the gas inflow can be depleted by sBHs if there is no feedback.  
This is the case irrespective of the SMBH mass and the gas inflow rate, as shown in 
Fig.~\ref{fig:dep_m}, which compares the depletion rate by sBHs and the inflow rate in the AGN disk. 
In this case, the growth of SMBHs may be dominated by the accretion of sBHs. 
This is in contradiction with Soltan's argument in which massive SMBHs have grown mainly via gas accretion during luminous AGN phases \citep[e.g.][]{Yu2002}.
We should note that this conclusion depends on $N_{\rm sBH,AGN}$, which cannot be well constrained due to uncertainties in the size of the AGN disk \citep[e.g.][]{Burtscher13,Stalevski2019}, 
in the stellar initial mass function in galactic centers \citep{Lu13}, and in
the anisotropy of the velocity dispersion of sBHs caused by vector resonant relaxation \citep{Szolgyen18}.

\noindent{(ii) "Over-growth Problem".}

\noindent In the Galactic center, a third object with a mass of $\gtrsim 100\,\Msun$ is prohibited within $R_{\rm sBH}\sim 10^{-3}$--$10^{-2}\,{\rm pc}$ from Sgr~A* \citep{Naoz2020,Gravity2020}. 
If sBHs efficiently grew in a possible AGN disk around Sgr~A* in $\sim 10\,{\rm Myr}$ \citep{Su2010}, 
there would not be enough time for 
the grown sBHs 
($M_{\rm sBH}\sim 10^2$--$10^3\,\Msun$) 
to migrate and 
merge to
Sgr~A* in the quiescent phase \citep[e.g.][]{Merritt2010}. 
Hence, the evolution of sBHs without feedback (Fig.~\ref{fig:r_m_noreg}) is likely inconsistent with observations of stellar orbital dynamics in the Galactic center.

To resolve these possible problems and contradictions, there ought to be some mechanism
to regulate the growth of sBHs in AGN disks. 
While there are established feedback processes regulating gas accretion onto sBHs, it is unclear whether these can sufficiently suppress the time-averaged accretion rate in these systems (Appendix~\ref{sec:feedback}), motivating the present study of a different mechanism.

\section{Regulation of gas accretion}

\begin{figure}
\begin{center}
\includegraphics[width=80mm]{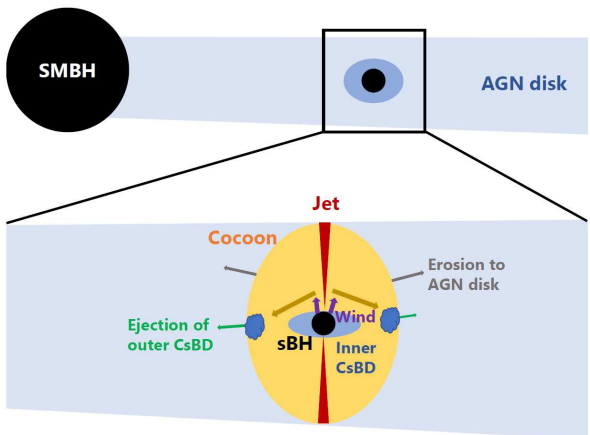}
\caption{
Schematic picture of the evolution of a cocoon produced around a jet. 
}
\label{fig:schematic}
\end{center}
\end{figure}

\subsection{Overview of the regulation process}

We begin with an overview of our model, in which gas accretion onto an sBH in an AGN disk is regulated by a cocoon produced around a jet. 
A schematic picture of the regulation is shown in Fig.~\ref{fig:schematic}. 
We propose that 
an accreting sBH embedded in an AGN disk experiences the following episodes. 

\begin{enumerate}

\item 
The jet evacuates a large hollow region ($r_{\rm dep}\sim 5\times 10^{15}\,{\rm cm}$; Eq.~\ref{eq:r_dep}) very quickly ($t_{\rm break}\sim 0.04~{\rm yr}$; Eq.~\ref{eq:t_break}). 

\item 
Inside this hollow region, the outer part of CsBD is blown away, and only a small inner portion of the CsBD survives (inside $r_{\rm tru}\sim 2\times 10^{14}\,{\rm cm}$; Eq.~\ref{eq:r_tru}).

\item
The inner CsBD accretes onto the sBH in a relatively short timescale ($t_{\rm cons}\sim 300~ {\rm yr}$; Eq.~\ref{eq:t_cons}), and accretion onto the sBH and jet production are quenched. 

\item 
The AGN gas refills the hollow cavity in a longer timescale ($t_{\rm re}\sim 5\times 10^{3}\,{\rm yr}$; Eq.~\ref{eq:t_re}). 
A jet then turns on again, and the cycle restarts from step 1. 
\end{enumerate} 

Due to the existence of the quenching phase 
and lower inflow rates onto the sBH in later episodes, 
the averaged accretion rate onto the sBH is reduced by $\sim 10-100$ 
($\S\,\ref{sec:acc_ave}$, Appendix~$\ref{sec:parameter_dep}$). 
Note that the relation $t_{\rm break}\ll t_{\rm cons}<t_{\rm re}$ implies the presence of recurring intermittent accretion, rather than a steady-state.

\subsection{Numerical choices}
\label{sec:numerical_choice}

We set $M_\mathrm{sBH,ini}=10\,\Msun$ 
since nuclear star clusters are rather metal rich \citep[e.g.][]{Do18,Schodel20} 
and remnant masses are roughly predicted to come around this value due to intense stellar mass loss \citep[e.g.][]{Belczynski10}. 
We assume that the initial radial position of the sBH is $R_{\rm sBH,ini}=1\,{\rm pc}$, 
the mass of the SMBH is $M_{\rm SMBH}=10^6\,\Msun$, 
the gas inflow rate from the outer boundary ($R_{\rm out}=5\,{\rm pc}$) of the AGN disk in units of the Eddington rate is ${\dot m}_{\rm SMBH}=0.1$,  
the angular momentum transfer parameter of the AGN disk is $m_{\rm AM}=0.15$ \citep{Thompson05}, 
the viscous parameter of the AGN disk is $\alpha_{\rm AGN}=0.1$ assuming partially ionized gas ($\gtrsim 10^3\,{\rm K}$) and 
that of the CsBD is $\alpha_{\rm CsBD}=0.3$ assuming fully ionized gas \citep{King07,Martin2019}, 
the jet energy conversion efficiency is $\eta_{\rm j}=0.5$ (Appendix~\ref{sec:jet_prod_app}), 
the opening angle of the injected jet is $\theta_0=0.2$, 
with the caveat that it is highly uncertain \citep[e.g.][]{Pushkarev2009,Hada2013,Hada2018,Berger2014}, 
and the reduction fraction of the accretion rate onto the sBH (${\dot M}_{\rm sBH}$) over the capture rate is 
$f_{\rm acc}={\dot M}_{\rm sBH}/{\dot M}_{\rm cap}=0.1$, 
which is assumed to be due to mass loss by wind/outflows and episodic accretion as described in $\S\,\ref{sec:resupply}$.

\subsection{Jet and cocoon}
\label{sec:jet_prod}

Here we briefly describe the properties of a jet and a surrounding cocoon. 
A relativistic jet is produced from a rotating sBH accreting at a super-Eddington rate (see Appendix~\ref{sec:jet_prod_app}) due to the BZ process with luminosity  
\begin{align}
\label{eq:l_j2}
{L}_{\rm j}&= \eta_{\rm j} {\dot M}_{\rm sBH}c^2 \nonumber\\
&\simeq 10^{42}~{\rm erg/s}~
\left(\frac{{\dot M}_{\rm cap}}{3\times 10^{-4}\,\Msun/{\rm yr}}\right)
\left(\frac{\eta_{\rm j} }{0.5}\right)
\left(\frac{f_{\rm acc}}{0.1}\right), 
\end{align}
where 
$c$ is the speed of light.

Expressing ${L}_{\rm j}$ in terms of $\theta_0$ and $\rho_{\rm AGN}$, 
the properties of the cocoon, the head velocity ($\beta_{\rm h}$), the cocoon velocity ($\beta_{\rm c}$), the cocoon opening angle ($\theta_{\rm c}$), and pressure ($P_{\rm c}$) are estimated as described in  Appendix~\ref{sec:cocoon_prop}.

\subsection{Evaporation of the CsBD}

To estimate the regulation of gas accretion as mediated by the evolution of the cocoon, 
it is crucial to understand how much gas in the AGN disk and the CsBD around the sBH is depleted after the cocoon propagation.  
Note that, in a constant density profile  $\rho \propto z^{0}$, the cocoon also propagates laterally along the $z=0$ plane \citep[e.g.][]{Wagner2011,Bromberg2014}, which likely results in an interaction with the CsBD. As the expanding cocoon hits the outer regions of the CsBD from above and below, the resulting shock heating can disrupt and unbind the outer regions of the disk. 
Due to the strong shocks, the pressure of the CsBD is roughly enhanced to the pressure of the cocoon ($P_{\rm c}$). 
The approximation of the uniform pressure within the shock downstream works for the shock propagation in the case that the internal pressure of the shocked medium dominates over any external pressure \citep[e.g.][]{Kompaneets1960}. 
If the sound speed of the shocked CsBD ($\sim [P_{\rm c}/\rho_{\rm CsBD}(r)]^{1/2}$) exceeds the escape velocity from the sBH at some distance $R$, 
the CsBD at distances $r>r_{\rm tru}$ is expected to be ejected \citep[e.g.][]{Perna2000}, where $\rho_{\rm CsBD}(r)$ is the density of the CsBD at $r$ before the shock heating and $r_{\rm tru}$ is the truncation radius. 
An estimate of $r_{\rm tru}$ can be made via the condition 
\begin{align}
\label{eq:p_ram_eq}
\rho_{\rm CsBD}(r_{\rm tru}) v_{\rm CsBD,Kep}(r_{\rm tru})^2
=
P_{\rm c}\,,
\end{align}
where $v_\mathrm{CsBD,Kep}(r)=\sqrt{GM_{\rm sBH}/r}$ is the Keplerian velocity of the CsBD at a distance $r$ from the sBH.

By solving Eq.~\eqref{eq:p_ram_eq} with $P_{\rm c}$ (Eq.~\ref{eq:p_c}), the truncation radius can be written as
\begin{align}
\label{eq:r_tru}
&r_{\rm tru}
\sim 2\times 10^{14}\,{\rm cm}~
\left(\frac{\rho_{\rm AGN}}{4\times 10^{17}~{\rm g/cm^3}}\right)^{-20/53}\nonumber\\
&\left(\frac{\beta_{\rm c}}{0.03}\right)^{-40/53}
\left(\frac{M_{\rm sBH}}{10\,\Msun}\right)^{31/53}
\left(\frac{{\dot M}_{\rm sBH}/\eta_{\rm rad}}{3\times 10^{-4}~\Msun/{\rm yr}}\right)^{8/53}\nonumber\\
&\left(\frac{\alpha_{\rm CsBD}}{0.3}\right)^{-14/53}
\left(\frac{\kappa_{\rm CsBD}}{2~{\rm cm^2/g}}\right)^{-6/53} 
\end{align} 
for gas pressure dominated regions of the CsBD 
(e.g. \citealt{Haiman2009}, black line of Fig.~\ref{fig:r_vari3}~a), where $\kappa_{\rm CsBD}$ is the opacity of the CsBD. 
The value of the opacity is consistently computed to satisfy the temperature and the density at $r_{\rm tru}$ for the standard disk model \citep{Shakura73}, using the opacity functions given by \citet{Bell94}. 
Within the photon trapping radius ($r_{\rm trap}\sim (3/2)r_{\rm g} [{\dot M}_{\rm sBH}/{\dot M}_{\rm Edd}(M_{\rm sBH})\eta_{\rm rad}]$ for an accretion disk, e.g. 
\citealt{Kato2008}, 
brown line in Fig.~\ref{fig:r_vari3}~a), 
we instead adopt the slim disk model, in which cooling is carried by advection \citep[e.g.][]{Kato2008}, 
where $r_{\rm g}$ is the gravitational radius. 
When $r_{\rm trap}>r_{\rm tru}$, the cocoon truncates a slim disk, which is found to be the case when a gap is assumed not to form (dashed lines in Fig.~\ref{fig:r_vari3}~a). 
The CsBD's mass within $r_{\rm tru}$ is 
\begin{align}
\label{eq:m_cbd}
&M_{\rm CsBD,tru} 
\sim \pi r_{\rm tru}^2 \Sigma_{\rm CsBD}(r_{\rm tru})\nonumber\\
&\sim 0.01\,{\rm \Msun}~
\left(\frac{\rho_{\rm AGN}}{4\times 10^{-17}\,{\rm g/cm^3}}\right)^{-28/53}
\left(\frac{\beta_{\rm c}}{0.03}\right)^{-56/53}\nonumber\\
&\left(\frac{M_{\rm sBH}}{10\,\Msun}\right)^{54/53}
\left(\frac{{\dot M}_{\rm sBH}/\eta_{\rm rad}}{3\times 10^{-4}\,\Msun/{\rm yr}}\right)^{43/53}\nonumber\\
&\left(\frac{\alpha_{\rm CsBD}}{0.3}\right)^{-62/53}
\left(\frac{\kappa_{\rm CsBD}}{2\,{\rm cm^2/g}}\right)^{-19/53}, 
\end{align}
where $\Sigma_{\rm CsBD}(r)$ is the surface density of the CsBD at $r$.

We set the consumption timescale of $M_{\rm CsBD,tru}$ to 
\begin{align}
\label{eq:t_cons}
t_{\rm cons}&=M_{\rm CsBD,tru}/{\dot M}_{\rm cap}\nonumber\\
&\sim 300\,{\rm yr}~
\left(\frac{M_{\rm CsBD,tru}}{0.01\,\Msun}\right)
\left(\frac{{\dot M}_{\rm cap}}{3\times 10^{-4}\,\Msun/{\rm yr}}\right)^{-1}
\end{align}
(black line in Fig.~\ref{fig:r_vari3}~b), 
which is comparable to the viscous timescale at this radius. 
Here, for simplicity, we assume that the accretion rate onto the sBH is unchanged after the cocoon ejects the outer regions of the CsBD. 
This assumption ignores the outward diffusion of the CsBD and the possible enhancement of the accretion rate due to heating by the shock, 
while it does not affect the estimate of the averaged accretion rate, as long as 
the fraction $\sim f_{\rm acc}$ of the bounded CsBD ($M_{\rm CsBD,tru}$) is accreted onto the sBH within the resupply timescale ($t_{\rm re}$, Eq.~\ref{eq:t_re}). 
For simplicity, the change of $f_{\rm acc}$ after the truncation is ignored in our estimate.

\subsection{Erosion of cocoon}

Next, we compute the radial extent from the sBH up to which the cocoon can propagate and eject gas in the AGN disk. 
This radial extent can be estimated by the Kompaneets method \citep{Kompaneets1960} for simple density profiles, by assuming a strong shock and uniform internal pressure within the shocked medium. 
\citet{Olano2009} showed that shocks can laterally propagate to $r\lesssim \pi H_{\rm AGN}$ for spherical explosions produced at $z=Z=r=0$ with a density profile of $\rho={\rm exp} (-|Z|/H_{\rm AGN})$.
Although the geometry of the cocoon and the density profiles in AGN disks may be different from those in \citet{Olano2009}, we approximately estimate the radial extend of the cocoon to be
\begin{align}
\label{eq:r_dep}
r_{\rm dep}&= f_{\rm ext} r_{\rm c} \nonumber\\
&\sim 5\times 10^{15}\,\mathrm{cm}~
\left(\frac{H_{\rm AGN}}{0.003\,{\rm pc}}\right)
\left(\frac{\theta_{\rm c}}{0.2}\right)
\left(\frac{f_{\rm ext}}{3}\right), ~~
\end{align}
(orange line in Fig.~\ref{fig:r_vari3}~a), where $f_{\rm ext}$ accounts for the geometrical and ambient-profile effects and  $r_{\rm c}\sim H_{\rm AGN}\theta_{\rm c}$ is the $r$-direction extent of the cocoon at the jet breakout, 
whose timescale is 
\begin{eqnarray}
\label{eq:t_break}
t_{\rm break}\sim H_{\rm AGN}/(\beta_{\rm h} c)
\sim 0.05\,{\rm yr}~
\left(\frac{H_{\rm AGN}}{0.003\,{\rm pc}}\right)
\left(\frac{{\tilde L}}{0.03}\right)^{-0.5}
\end{eqnarray}
(blue line in Fig.~\ref{fig:r_vari3}~b). 
After the jet breakout, the cocoon also escapes in the vertical direction on a timescale which is of the same order as $t_{\rm break}$.

\subsection{Resupply of gas}

\label{sec:resupply}

After the cocoon escapes vertically from the AGN disk plane, gas outside $r_{\rm dep}$ is resupplied to the sBH. 
The resupply timescale of gas to the CsBD is roughly given by 
\begin{align}
\label{eq:t_re}
t_{\rm re}
&\sim r_{\rm dep}/c_{\rm s, AGN}\nonumber\\
&\sim 5\times 10^3\,{\rm yr}~
\left(\frac{r_{\rm dep}}{6\times 10^{15}\,{\rm cm}}\right)
\left(\frac{c_{\rm s,AGN}}{0.4~{\rm km/s}}\right)^{-1} 
\end{align}
(orange line in Fig.~\ref{fig:r_vari3}b, e.g. \citealt[][]{Tanigawa2012}). 
Gas inflow onto sBHs preferentially proceeds from the direction perpendicular to the AGN disk plane, 
as found in high-resolution three-dimensional hydrodynamical simulations \citep{Ayliffe2009,Tanigawa2012,Szulagyi2021}. 
Such inflow structure enables multiple episodes of cocoon evolution by ensuring interactions of the jets and inflowing gas.

In the following we make a rough estimate of the accretion rate onto the sBH after the resupply. 
According to the hydrodynamical simulation for gas flow onto a circum-planetary disk by \citet{Tanigawa2012}, 
gas captured by the sBH vertically falls onto the CsBD with some angular momentum, 
and subsequently it quickly circularizes at the corresponding Keplerian radius. 
The total mass 
accretion 
rate onto the CsBD 
inside some distance $r$ 
roughly follows 
\begin{align}
\label{eq:md_cbd_kep}
{\dot M}_{\rm CsBD,Kep}(<r)
\sim
\left\{
\begin{array}{l}
{\dot M}_{\rm cap} (r/r_{\rm CsBD,out})^{0.5}\\
~~~~~~~~~~\mathrm{for}~r\lesssim r_{\rm CsBD,out}, \\
{\dot M}_{\rm cap} 
~~~\mathrm{otherwise} 
\end{array}
\right.
\end{align}
(Fig.~14 of \citealt{Tanigawa2012}), 
where $r_{\rm CsBD,out}$ is the outer radius of the CsBD. 
We parameterize it as $r_{\rm CsBD,out}\sim f_{\rm circ}\,r_{\rm Hill}$, where $f_{\rm circ}=0.1$--$0.4$ is inferred from simulations of circum-planetary disks \citep{Ayliffe2009,Martin2011,Tanigawa2012}. 
The circularized gas 
inflows toward 
the sBH on the viscous timescale. Equating the viscous timescale to the resupply timescale, we can obtain the critical radius, $r_{\rm vis}$, beyond which the bulk of the gas cannot accrete within the resupply timescale. 
In the fiducial model, $r_{\rm vis}\sim 0.01\,r_{\rm Hill}$. 
Then, the gas inflow rate to the sBH within $r_{\rm vis}$ is reduced to 
\begin{align}
{\dot M}_{\rm in,vis}\sim {\dot M}_{\rm CsBD,Kep}(<r_{\rm vis})
\sim 0.3~{\dot M}_{\rm cap}~\left(\frac{f_{\rm circ}}{0.1}\right)^{-1/2}. 
\end{align}
Note that the reconstituted CsBD can extend beyond $r_{\rm vis}$, but its outer regions, beyond this radius, will not have reached equilibrium 
during the disk accretion episodes. 
We assume that the reduction of the 
inflow 
rate of the CsBD in later episodes is effectively taken into account in $f_{\rm acc}$ ($\S\,\ref{sec:numerical_choice}$).

Once gas resupplies 
and accretion begins, 
we assume that jets are launched soon (on a timescale of $\lesssim 1~{\rm s}$, e.g. \citealt{Narayan2021}, see also Appendix~\ref{sec:jet_prod_app}).

\subsection{Time-averaged accretion rate}

\label{sec:acc_ave}

The average accretion rate in  units of the Eddington rate is estimated by multiplying the active fraction ($t_{\rm cons}/t_{\rm re}$, Fig.~\ref{fig:r_vari3}~b) as 
\begin{align}
\label{eq:md_bhl_edd_ave}
&{\dot m}_{\rm sBH,ave}
=\frac{{\dot M}_{\rm sBH}}{{\dot M}_{\rm Edd}(M_{\rm sBH})}
\frac{t_{\rm cons}}{t_{\rm re}}\nonumber\\
&\sim 10~
\left(\frac{\rho_{\rm AGN}}{4\times 10^{-17}\,{\rm g/cm^3}}\right)^{-28/53}
\left(\frac{{\dot M}_{\rm sBH}}{3\times 10^{-5}\,{\Msun/{\rm yr}}}\right)^{43/53}\nonumber\\
&\left(\frac{\beta_{\rm c}}{0.03}\right)^{-56/53}
\left(\frac{M_{\rm sBH}}{10\,\Msun}\right)^{1/53}
\left(\frac{\eta_{\rm rad}}{0.1}\right)^{10/53}
\left(\frac{f_{\rm acc}}{0.1}\right)
\nonumber\\
&\left(\frac{\alpha_{\rm CsBD}}{0.3}\right)^{-62/53}
\left(\frac{\kappa_{\rm CsBD}}{2~{\rm cm^2/g}}\right)^{-19/53}
\left(\frac{\theta_{\rm c}}{0.2}\right)^{-1}\nonumber\\
&\left(\frac{H_{\rm AGN}}{0.003\,{\rm pc}}\right)^{-1}
\left(\frac{f_{\rm ext}}{3}\right)^{-1}
\left(\frac{c_{\rm s,AGN}}{0.4~{\rm km/s}}\right). 
\end{align}
In the fiducial setting, the accretion rate is reduced by a factor of $\sim 60$, ($t_{\rm re}/t_{\rm cons}\sim 20$ and a further reduction by $\sim 3$ due to the depletion of the inflow rate onto the sBH described in $\S\,\ref{sec:resupply}$) 
compared to that without the 
JFM 
(another factor of $\sim3$ is assumed to be reduced by wind/outflow feedback), 
and the accretion rate becomes moderately super-Eddington by a factor ${\dot M}_{\rm sBH}/{\dot M}_{\rm Edd}(M_{\rm sBH})\sim 10$ (Eq.~\ref{eq:md_bhl_edd_ave}, black line in Fig.~\ref{fig:r_vari3}~c).

\begin{figure}
\begin{center}
\includegraphics[width=80mm]{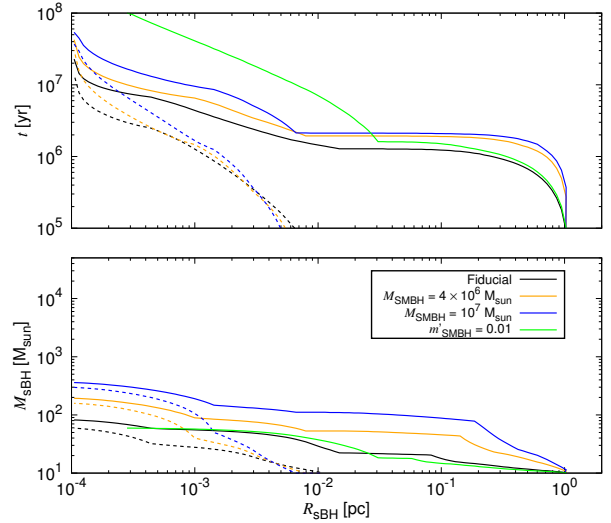}
\caption{
Same as 
Fig.~\ref{fig:r_m_noreg}, but with the 
JFM. 
As the comparison of the two figures shows, the final sBH masses are reduced by approximately 
one order 
of magnitude.
}
\label{fig:r_m}
\end{center}
\end{figure}

\section{Discussion}

\label{sec:discussion}

We now discuss the influence of the 
JFM 
on 
the over-growth and 
depletion 
problems highlighted in $\S\,\ref{sec:rapid_growth}$.  
Fig.~\ref{fig:r_m} shows the evolution of the sBH in an AGN disk calculated in the same way as in Fig.~\ref{fig:r_m_noreg}, but accounting for 
the JFM. 
With feedback, the growth of sBHs is suppressed compared to that without feedback (lower panels of Figs.~\ref{fig:r_m} and \ref{fig:r_m_noreg}). 
Although the average accretion rate shown in Fig.\ref{fig:r_vari3}~c is still significant, especially in inner regions of $R_{\rm sBH}\lesssim 0.1\,{\rm pc}$, sBHs can open a gap at $R_{\rm sBH}\lesssim 0.01\,{\rm pc}$ and the accretion rate is further reduced by a factor of $\sim 100\,(M_{\rm sBH}/100\,\Msun)^2$ for massive sBHs 
(Figs.~\ref{fig:r_m_noreg}, \ref{fig:r_m}, and \ref{fig:m_dep}) 
compared to the black line of Fig.~\ref{fig:r_vari3}~c. 
Here, in gap regions, the density of the AGN disk becomes lower, reducing the cocoon pressure and weakening the 
JFM. 
On the other hand, if a gap does not form as suggested by different disk models, the regulation by the cocoon 
is 
efficient 
in the inner regions of $R\lesssim 0.01\,{\rm pc}$ 
as a low density-inner slim CsBD is efficiently ejected 
(the dashed lines of Fig.~\ref{fig:r_vari3}~c) 
by the high-pressure cocoon in the high-density AGN disk. 
Due to the suppression of accretion, 
sBHs migrate to $R_{\rm sBH}\sim 10^{-4}\,{\rm pc}$ almost without passing through the regions prohibited 
by the S2 orbits 
($\S\,\ref{sec:rapid_growth}$, see also \citealt{Gravity2020}). Hence, the over-growth problem can be avoided.

Next, we consider the depletion problem.  
With 
\begin{align}
M_{\rm sBH,fin}\sim 100\,\Msun~ 
\left(\frac{M_{\rm SMBH}}{10^6\,\Msun}\right)
\left(\frac{{\dot m}_{\rm SMBH}}{0.1}\right)^{1/2} 
\end{align}
at $t_{\rm AGN}=10$--$100\,{\rm Myr}$ 
roughly derived from Fig.~\ref{fig:r_m} 
and Eq.~\eqref{eq:n_agn_sbh}, 
the depletion rate is 
\begin{align}
&{\dot M}_{\rm sBH,tot}\sim M_{\rm sBH,fin} N_{\rm sBH,AGN}/t_{\rm AGN} \nonumber\\
&\sim 3\times 10^{-3} - 3\times 10^{-4}\,\Msun /{\rm yr}~
\left(\frac{M_{\rm SMBH}}{10^6\,\Msun}\right)
\left(\frac{{\dot m}_{\rm SMBH}}{0.1}\right),  
\end{align}
which is smaller than the inflow rate of 
\begin{align}
{\dot M}_{\rm SMBH,in}\simeq 2\times 10^{-3}\,\Msun/{\rm yr}~ 
\left(\frac{M_{\rm SMBH}}{10^6\,\Msun}\right)
\left(\frac{{\dot m}_{\rm SMBH}}{0.1}\right)
\end{align}
in later phases of $t_{\rm AGN}\gtrsim$ a few $10~{\rm Myr}$ 
with the same dependence (Fig.~\ref{fig:dep_m}). 
Note that the ejection mass of gas orbiting above the sBH (with the density $\rho_{\rm AGN}$) 
is calculated to be smaller in most cases than the mass accreted by the sBH (see Appendix~\ref{sec:parameter_dep}), so that this ejected gas does not exacerbate the depletion problem. 
Also, the ejected CsBDs can be recaptured by the AGN disk. 
The radius within which the sonic velocity after shock heating by the cocoon is lower than the escape velocity from the SMBH ($r_{\rm esc}$) exceeds the CsBD size ($r_{\rm CsBD,out}$) 
for 
\begin{align}
R_{\rm sBH}\gtrsim 0.1\,{\rm pc} 
\left(\frac{f_{\rm circ}}{0.1}\right)
\left(\frac{M_{\rm SMBH}}{10^6\,\Msun}\right)
\left(\frac{{\dot m}_{\rm SMBH}}{0.1}\right)
\end{align}
in the fiducial settings. 
For $M_{\rm SMBH}\sim10^6\,\Msun$, the condition ${\dot M}_{\rm sBH,tot}<{\dot M}_{\rm inflow}$
is marginally satisfied, 
while for $M_{\rm SMBH}\lesssim 10^5\,\Msun$ 
inflowing gas may be depleted by the 
JFM. 
as the ejected CsBDs often escape. 
Interestingly, 
the regulation by the 
JFM 
may be consistent with observations indicating that high-Eddington ratio AGNs are rare for $M_{\rm SMBH}\lesssim 10^5\,\Msun$ \citep{Greene07}. 
In either case, the depletion problem can be alleviated by the 
JFM.

Our results depend on several assumptions. 
First, we assumed the fraction of the outer radius of the CsBD over 
the Hill radius ($f_{\rm circ}=r_{\rm CsBD,out}/r_{\rm Hill}$), ${\dot M}_{\rm CsBD,Kep}(<r)$, and we relied on a gas inflow geometry derived in studies of proto-planetary systems. 
Second, $f_{\rm ext}$ and $r_{\rm tru}$ are roughly determined by assuming uniform pressure inside the shocked medium. 
Third, hydrodynamical jets are assumed to be produced once super-Eddington accretion onto a spinning sBH is realized. 
The regulation of accretion by the cocoon can be drastically influenced by the failure of these assumptions. 
If it fails, 
a possible alternative process for regulation would be that of a significant outflow compared to inflow on the sBHs -- i.e. suggesting that highly super-Eddington accretion is self-regulated. 
Thus AGN disks can be good experimental environments to understand the processes regulating highly super-Eddington accretion flows.

\section{Conclusions}

In this paper, 
for the first time, we considered the jet feedback mechanism 
for regulating gas accretion onto sBHs in AGN disks, 
in which 
a cocoon generated around a jet 
launched by an accreting sBH due to the BZ effect can regulate the average accretion onto the sBH over the AGN lifetime.
Additionally, we find that a similar regulation can occur in the outer regions of the AGN disk due to winds from the embedded sBHs  (Appendix~\ref{sec:wind_feedback}).
Our main results are summarized as follows:

\begin{enumerate}

\item 
We highlight two problems for gas accretion onto sBHs in AGN disks. One is the over-growth of sBHs and the other is the depletion of gas inflow in AGN disks by the population of sBHs in the disk, starving the central SMBH.

\item Due to the cocoon evolution, the accretion rate onto sBHs embedded in AGN disks is reduced by a factor of $\sim 10$--$100$, depending on parameters. 

\item The problems of the depletion of gas inflow and 
over-growth of massive sBHs can be avoided by considering the 
jet feedback mechanism.

\item Efficient gas ejection and depletion of gas inflow in AGN disks around less massive SMBHs of $\lesssim 10^5\,\Msun$ expected by the 
jet feedback mechanism 
may explain the dearth of quasars with high accretion rates for such SMBHs. 

\end{enumerate}

These findings suggest that cocoons play a significant role in regulating accretion in AGN disks and the evolution of embedded compact objects. 
Finally, we note that shocks produced by winds from CsBDs also play a role in regulating accretion
in the outer regions $R\gtrsim 10^{-1}~{\rm pc}$ of the AGN disk.

\acknowledgments

The authors thank 
Sunmyon Chon, 
Kazuyuki Sugimura, 
Kazuyuki Omukai, 
Hiromi Saida, and 
Yohsuke Takamori 
for meaningful discussions. 
This work was financially supported 
by Japan Society for the Promotion of Science (JSPS) KAKENHI 
Grant Number JP21J00794 (HTagawa), JP19J00198 (SSK), 
and 18H05438 (HTanaka). 
ZH was supported by NASA grant NNX15AB19G and NSF grants AST-2006176 and AST-1715661.
RP acknowledges support by NSF award AST-2006839 and from NASA (Fermi) award 80NSSC20K1570.

\appendix

\section{Mechanisms}

Here we provide further details on the mechanisms related to the regulation of gas accretion by the cocoon. 

\subsection{Jet production}
\label{sec:jet_prod_app}

We outline how a jet is launched from an accreting sBH embedded in an AGN disk. In the AGN disk, an sBH is surrounded by a CsBD (Fig.~\ref{fig:schematic}). 
When the CsBD is advection dominated, as expected here, a magnetically dominated state can be realized \citep[e.g.][]{Meier2001,Kimura2021_BBH_PeV} due to the accumulation of the magnetic flux in the vicinity of the sBH \citep{Cao2011}. Even if the magnetic flux is initially weak, the outflow from the disk converts the toroidal magnetic field generated by the shear motion into a poloidal field \citep{Liska2020}. Such advection-dominated flows are expected for super-Eddington accretion rates \citep{Abramowicz1988} or low accretion rates of $\dot{M}_{\rm sBH}\lesssim 0.01\,{{\dot M}_{\rm Edd}}(M_{\rm sBH})$ \citep[e.g.][]{Narayan1994,Blandford1999}. 
In these cases, the jets from  spinning BHs can be launched through the BZ process \citep{Blandford1977}.
We assume that the luminosity of the relativistic jet is proportional to the mass accretion rate onto the sBH as 
\begin{align}
\label{eq:l_j1}
L_{\rm j}= \eta_{\rm j} {\dot M}_{\rm sBH}c^2 
\end{align}
\citep{Blandford1977}. 
Note that the jet is assumed to be launched once accretion begins due to efficient magnification of the magnetic field during gas inflow \citep{Cao2011}. On the other hand, if the launch of the jet is delayed by a timescale longer than the consumption timescale $t_{\rm cons}$, then the average accretion rate onto the sBH is enhanced by the ratio of those timescales compared to Eq.~\eqref{eq:md_bhl_edd_ave}. 
Here, $\eta_{\rm j}$ is approximated as 
$\eta_{\rm j}\sim a_{\rm sBH}^2$ for a magnetically dominated state \citep[e.g.][]{Tchekhovskoy2010,Tchekhovskoy2011,Narayan2021}. 
In the fiducial model, we adopt the conversion efficiency to jet to be $\eta_{\rm j}=0.5$, assuming a merger remnant with a dimensionless spin of $a_{\rm sBH}\sim 0.7$ 
(e.g. \citealt{Buonanno08}, a different value for $\eta_{\rm j}$ is investigated in Appendix~\ref{sec:parameter_dep_final_mass}). 
We ignore the spin evolution of sBHs by gas accretion and production of the BZ jet \citep{Narayan2021}, which affects $\eta_{\rm j}$ and accordingly the regulation rate of accretion by the cocoon. 
In our model, a jet is assumed to be launched whenever ${\dot M}_{\rm sBH}>{\dot M}_{\rm Edd}$. 

\subsection{Cocoon propagation}

\label{sec:cocoon_prop}

We describe the properties of the cocoon evolving in the AGN disk following the formalism of \citet{Bromberg2011}, who investigate hydrodynamic jets. 
Although a BZ jet is considered to be an initially magnetized jet, we adopt the formulae for hydrodynamic jets for simplicity.

When the jet collides with the AGN disk orbiting above an sBH, two shocks forms: a forward shock propagating in the AGN disk and a reverse shock in the jet. The region sandwiched by the two shocks is called the jet head. 
The head velocity is estimated as 
\begin{align}
\label{eq:beta_h}
\beta_{\rm h}\sim 
\left\{
\begin{array}{l}
{\tilde L}^{1/2}
\qquad~~~~~~\mathrm{for}~{\tilde L}<1, \\
1
\qquad~~~~~~~~~~\mathrm{otherwise}, 
\end{array}
\right.
\end{align}
where 
\begin{align}
\label{eq:l_tilde}
{\tilde L}
\sim \left\{
\begin{array}{l}
\left(\frac{L_{\rm j}}{\rho_{\rm AGN} t_{\rm j}^2 \theta_0^4 c^5}\right)^{2/5}
\qquad~~~\mathrm{for}~{\tilde L}<\theta_0^{-4/3}, \\
\frac{L_{\rm j}}{\rho_{\rm AGN} t_{\rm j}^2 \theta_0^2 c^5}
\qquad~~~~~~~~~~\mathrm{otherwise}, 
\end{array}
\right.
\end{align}
is the ratio between the energy density of the jet and the rest-mass energy density of the surrounding medium at the location of the head, 
where 
$t_{\rm j}$ is the time since the jet is launched.

If the energy stored in the cocoon ($E_{\rm c}$) is uniformly distributed within the cocoon's volume ($V_{\rm c}$), 
the pressure of the cocoon is 
\begin{align}\label{eq:p_c}
P_{\rm c}=E_{\rm c}/3V_{\rm c}
\sim \left\{
\begin{array}{l}
{\tilde L}\theta_0^2 \rho_{\rm AGN} c^2
\qquad~~~~~~\mathrm{for}~{\tilde L}<\theta_0^{-4/3}, \\
{\tilde L}^{1/4}\theta_0 \rho_{\rm AGN} c^2
\qquad~~~\mathrm{for}~\theta_0^{-4/3}<{\tilde L}<4\Gamma_{j}^4, \\
\Gamma_{\rm j} \theta_0 \rho_{\rm AGN} c^2
\qquad~~~~~~\mathrm{otherwise}. 
\end{array}
\right.
\end{align}
where $\Gamma_{\rm j}$ is the Lorentz factor just below the head. 
In Eq.~\eqref{eq:p_c}, it is assumed that the radiation pressure dominates the gas pressure, which is valid as long as the cocoon is optically thick, 
since the cocoon has a high temperature, and the radiation and gas pressures evolve following a similar scaling in the adiabatic phase \citep[e.g.][]{Kashiyama2013}. 
Note that in the optically thin regime (dashed orange line in Fig.~\ref{fig:wind}~b), 
the Coulomb timescale (dashed black line in Fig.~\ref{fig:wind}~c) is longer than the breakout timescale (blue line in Fig.~\ref{fig:r_vari3}~b), implying that 
cooling is negligible for the cocoon evolution. 
In such regions, the gas pressure likely dominates the radiation pressure, and the cocoon pressure is given by $P_{\rm c}=2E_{\rm c}/3V_{\rm c}$. For instance, we calculated that the enhancement of the cocoon pressure by a factor of $2$ reduces the final sBH mass (e.g. Fig.~\ref{fig:r_m}) by a factor of $\sim 1.3$ in the fiducial model. 

The lateral expansion velocity of the cocoon is 
\begin{align}\label{eq:beta_c}
\beta_{\rm c}
=\sqrt{P_{\rm c}/\rho_{\rm AGN} c^2}
\sim \left\{
\begin{array}{l}
{\tilde L}^{1/2}\theta_0
\qquad~~~~~~\mathrm{for}~{\tilde L}<\theta_0^{-4/3}, \\
{\tilde L}^{1/8}\theta_0^{1/2}
\qquad~~~\mathrm{for}~\theta_0^{-4/3}<{\tilde L}<4\theta_{0}^{-4}, \\
1
\qquad~~~~~~\mathrm{otherwise}, 
\end{array}
\right. 
\end{align}
and the opening angle of the cocoon is 
\begin{align}\label{eq:theta_c}
\theta_{\rm c} \sim \beta_{\rm c}/\beta_{\rm h}
\sim \left\{
\begin{array}{l}
\theta_0
\qquad~~~~~~\mathrm{for}~{\tilde L}<1, \\
{\tilde L}^{1/2}\theta_0
\qquad~~~\mathrm{for}~1<{\tilde L}<\theta_0^{-4/3}, \\
{\tilde L}^{1/8}\theta_0^{1/2}
\qquad~~~\mathrm{for}~\theta_0^{-4/3}<{\tilde L}<4\theta_{0}^{-4}, \\
1
\qquad~~~~~~\mathrm{otherwise}. 
\end{array}
\right.
\end{align}

At breakout of the cocoon from the AGN disk ($t_{\rm j}= \frac{H_{\rm AGN}}{\beta_{\rm h} c}$), 
from Eqs.~\eqref{eq:beta_h} and \eqref{eq:l_tilde} we obtain
\begin{align}
\label{eq:l_tilde_bo}
{\tilde L}
\sim \left\{
\begin{array}{l}
\left(\frac{L_{\rm j}^2}{\rho_{\rm AGN}^2 \theta_0^8 c^6 H_{\rm AGN}^4}\right)^{1/3}
\qquad~~~\mathrm{for}~{\tilde L}<1, \\
\left(\frac{L_{\rm j}}{\rho_{\rm AGN} \theta_0^4 c^3 H_{\rm AGN}^2}\right)^{2/5}
\qquad~~~\mathrm{for}~1<{\tilde L}<\theta_0^{-4/3}, \\
\frac{L_{\rm j}}{\rho_{\rm AGN} \theta_0^2 c^3 H_{\rm AGN}^2}
\qquad~~~~~~~~~~\mathrm{otherwise}. 
\end{array}
\right.
\end{align}
With the fiducial values at $R=1\,{\rm pc}$ (Table~\ref{table:parameter_fiducial}), this becomes
\begin{align}
\label{eq:l_tilde_bo_values}
{\tilde L}
\simeq 0.03~
\left(\frac{L_{\rm j}}{10^{42}\,{\rm erg/s}}\right)^{2/3}
\left(\frac{\rho_{\rm AGN}}{4\times 10^{-17}\,{\rm g/cm^3}}\right)^{-2/3}
\left(\frac{H_{\rm AGN}}{0.003~{\rm pc}}\right)^{-4/3}
\left(\frac{\theta_{0}}{0.2}\right)^{-8/3}, 
\end{align}
and accordingly, 
$\beta_{\rm h}\simeq 0.2({\tilde L}/0.03)^{1/2}$, 
$\beta_{\rm c}\simeq 0.04({\tilde L}/0.03)^{1/2}(\theta_0/0.2)$, 
and $\theta_{\rm c}\sim0.2(\theta_0/0.2)$. 
Thus, the properties of the cocoon are estimated by 
$L_{\rm j}$, 
$\rho_{\rm AGN}$, 
$H_{\rm AGN}$, and 
$\theta_{0}$.

\subsection{Evolution of black hole mass and location}
\label{sec:bh_growth}

The time evolution of the mass 
for an sBH initially at the position 
$R_{\rm sBH}(t=0)=R_{\rm ini}$ and mass $M_{\rm sBH}(t=0)=M_{\rm sBH,ini}$ without feedback is calculated as 
\begin{align}
\label{eq:macc_tagn}
M_{\rm sBH}(t)
=M_{\rm sBH,ini}+\int_{0}^{t} {\dot M}_{\rm cap}(R_{\rm sBH}(t'),M_{\rm sBH}(t')) dt', 
\end{align} 
while the sBH time-dependent position within the AGN disk is calculated as 
\begin{align}
\label{eq:r_t_int}
R_{\rm sBH}(t)
= R_{\rm ini}-\int_0^t dt' R_{\rm sBH}(t')/t_{\rm mig}(R(t'),M_{\rm sBH}(t')) ,
\end{align} 
where $t_{\rm mig}$ is the migration timescale, 
for which we adopt Eq.~\eqref{eq:typeI_II} in 
the next section. 
Figs.~\ref{fig:r_m_noreg}, \ref{fig:r_m}, \ref{fig:m_dep}, and \ref{fig:meje_dep} are calculated using Eqs.~\ref{eq:macc_tagn} and \ref{eq:r_t_int}.

\subsection{Migration}
\label{sec:migration}

To calculate how sBHs migrate radially towards the central SMBH (Eq.~\ref{eq:r_t_int}, Figs.~\ref{fig:r_m_noreg} and \ref{fig:r_m}~a), 
we adopt the formulae for the timescale of migration of sBHs as 
\begin{equation}
\label{eq:typeI_II}
t_\mathrm{I,II}=\frac{\Sigma_\mathrm{disk}}{\Sigma_\mathrm{disk,min}}t_\mathrm{I}
\end{equation}
\citep{Duffell14,Kanagawa18}, where 
$t_\mathrm{I}$ is the type I migration timescale, given by
\begin{eqnarray}
\label{eq:typeI}
t_\mathrm{I}\simeq \frac{1}{2f_\mathrm{mig}}
\left(\frac{M_\mathrm{SMBH}}{M_\mathrm{sBH}}\right)
\left(\frac{M_\mathrm{SMBH}}{\Sigma_\mathrm{disk} R_{\rm sBH}^2}\right)
\left(\frac{H_\mathrm{AGN}}{R_{\rm sBH}}\right)^{2}
\Omega_{\rm Kep}^{-1},
\end{eqnarray}
\citep[e.g.][]{Ward97,Tanaka02,Paardekooper10,Baruteau11}, 
where $\Sigma_{\rm disk}$ is the surface density of the AGN disk, 
$\Omega_{\rm Kep}$ is the Keplerian angular velocity around the SMBH, and
$f_\mathrm{mig}$ is a dimensionless factor depending on the local temperature and density profiles \citep[see][]{Paardekooper10,Baruteau11}. 
We set $f_\mathrm{mig}=2$ in the fiducial model, which is the typical value numerically found by \citet{Kanagawa18}. 
\citet{Fung14} and \citet{Kanagawa15} estimated that due to the gravitational torque from a migrator, the surface density in the disk annulus at the migrator's orbit is reduced to 
\begin{equation}
\label{eq:typeI_II_Sigma}
\Sigma_\mathrm{disk,min}=\frac{\Sigma_\mathrm{disk}}{1+0.04K}, 
\end{equation} 
where 
\begin{equation}
\label{eq:K_gap}
K=(M_\mathrm{sBH}/M_\mathrm{SMBH})^2(H_\mathrm{AGN}/R_{\rm sBH})^{-5}\alpha_\mathrm{eff}^{-1}, 
\end{equation} 
and $\alpha_\mathrm{eff}=\nu /(c_{\rm s,AGN} H_\mathrm{AGN})$ is the effective $\alpha$ viscosity parameter. 
In the outer regions of AGN disks where angular momentum transfer is presumed to be driven by torques from stellar bars, spiral waves, or large-scale magnetic stresses, this parameter is given by
$\alpha_\mathrm{eff}=m_{\rm AM} R/H_{\rm AGN}$.

For reference, it would be useful to present the typical sBH mass above which migration becomes efficient. 
If we assume ${\dot M}_{\rm sBH,ave}={\dot M}_{\rm cap} f_{\rm acc,eff}$ with a constant $f_{\rm acc,eff}$, 
since the capture timescale ($t_{\rm acc}=M_{\rm sBH}/{\dot M}_{\rm sBH,ave}\propto M_{\rm sBH}^{1/3}$) weakly depends on $M_{\rm sBH}$ compared to the migration timescale ($t_{\rm I}\propto M_{\rm sBH}$) in the fiducial model, 
there is a critical mass above which the sBH efficiently migrates in cases without gaps \citep{Tanaka2020}. 
By equating Eqs.~\eqref{eq:md_bondi} and \eqref{eq:typeI}, the critical mass at which $t_{\rm acc}=t_{\rm I}$ is 
\begin{eqnarray}
\label{eq:m_crit}
M_{\rm crit}=M_{\rm SMBH} [f_{\rm acc,eff}(1/3)^{2/3}(f_{\rm c}/4f_{\rm mig})(H_\mathrm{AGN}/R_{\rm sBH})^2]^{3/4}\nonumber\\
\sim 1.1\times 10^{-4}~M_{\rm SMBH} 
\left(\frac{f_{\rm acc,eff}}{1}\right)^{3/4} 
\left(\frac{f_{\rm c}}{10}\right)^{3/4} 
\left(\frac{f_{\rm mig}}{2}\right)^{-3/4} 
\left(\frac{H_\mathrm{AGN}/R_{\rm sBH}}{0.003}\right)^{3/2}. 
\end{eqnarray} 
Since $f_{\rm acc,eff}$ is $1$ and $\sim 1/60$ without and with the 
JFM 
at $R_{\rm sBH}=1~{\rm pc}$, respectively, 
the critical mass is $\sim 100\,\Msun~(M_{\rm SMBH}/10^6\,\Msun)$ and $\sim 5\,\Msun~(M_{\rm SMBH}/10^6\,\Msun)$, which are roughly consistent with the turning points of the lines in the lower panels of Figs.~\ref{fig:r_m_noreg} and \ref{fig:r_m}. 
It is found from Eq.~\eqref{eq:m_crit} that as $H_\mathrm{AGN}/R_{\rm sBH}$ increases, the growth of the sBH becomes efficient compared to the migration. Then, stronger feedback (lower $f_{\rm acc,eff}$) presumably becomes required to resolve the over-growth problem.

\section{Feedback}

\subsection{Influence on gas inflow}
\label{sec:feedback}

While sBHs are accreting, 
several kinds of feedback possibly affect the gas dynamics: 
(i) radiation pressure inside the CsBD
and 
wind/outflow from a thick disk or line driven wind; 
(ii) thermal pressure of ionized gas; (iii) radiation pressure on dust by infrared photons; 
(iv) dynamical instabilities; 
(v) gap formation. 
We evaluate whether these processes can regulate gas accretion onto sBHs in AGN disks.

(i) When the accretion rate exceeds the Eddington rate, the radiation pressure on ionized gas exceeds the gravity from the sBH, which can regulate gas accretion 
(while accretion might be promoted by the positive JFM for nearly spherical flows, \citealt{Chamandy2018}). 
However, in such highly accreting cases, cooling of the CsBD in the vicinity of the sBH is dominated by advection 
(neglecting neutrino cooling, which is efficient only for extremely rapid accretion with rates of $\gtrsim 10^{4}~\Msun/{\rm yr}$,
e.g. \citealt[][]{DiMatteo2002,Janiuk2004,Kohri2005,Chen2007_neutrino}). 
Then, photons can be trapped into the accretion flow and be accreted onto the sBH or escape in the vertical direction with respect to the CsBD 
\citep[e.g.][]{Jiang+2014,Sadowski+2015}. 
In such highly accreting flows, 
wind/outflow feedback also pushes back the accreting gas. 
If the wind is isotropically emitted, it significantly regulates accretion \citep{Wang2021_TZW} and gas dynamical friction \citep{Gruzinov2020}. 
On the other hand, if it is anisotropic, 
as predicted by numerical simulations \citep{Jiao2015,Kitaki2021}, 
the feedback does not efficiently regulate either accretion \citep{Takeo2020} or dynamical friction \citep{Li2020_gasDF}. 
Also, when the trapping radius is smaller than the circularization radius, the mass ratio of outflows to inflows is modest \citep{Kitaki2021}.

If the outflow rate is much higher than the accretion rate onto the sBH due to wind (e.g. $f_{\rm acc}\lesssim 0.01$), 
the over-growth problem can be avoided, while the depletion problem is not resolved unless the wind is decelerated. 
This is because the wind velocity at launch 
($\S\,\ref{sec:inf_wind}$) exceeds 
the escape velocity from the SMBH
($v_{\rm esc}=(2G M_{\rm SMBH}/R_{\rm sBH})^{1/2}$)
for 
\begin{eqnarray}
R_{\rm sBH}\gtrsim 3\times 10^{-4}~{\rm pc}~
\left(\frac{M_{\rm SMBH}}{10^6\,\Msun}\right)~
\left(\frac{{\dot m}_{\rm sBH}}{10^5}\right). 
\end{eqnarray}
This condition is satisfied most of the time during the evolution of sBHs in AGN disks, 
and the wind does not decelerate when it is driven continuously and gas above the sBH is ejected. 
As a result, 
even though the significant outflows in CsBDs can prevent the sBHs from growing overly massive, the AGN disk is still depleted, because the outflowing material is unbound and ejected, rather than retained in the AGN disk. 
On the other hand, the feedback by thermalized winds remains to be investigated in detail. 
In Appendix~$\ref{sec:wind_feedback}$, we estimate regulation of accretion onto the sBH by these thermal winds.

(ii) The thermal pressure of HII regions formed by radiation from the accreting sBHs 
can push back the gas accreting onto the sBHs themselves, and thus reduce the accretion rate 
\citep{Milosavljevic2009,Park2011,Park2012}. 
On the other hand, when the size of HII regions around the sBHs (Str{\" o}mgren radius, $r_{\rm HII}$) is less than $r_{\rm BHL}$, the dynamics of accreting gas is less influenced by the pressure \citep{Inayoshi2016_HyperEdd,Toyouchi20}. 
The condition $r_{\rm HII}<r_{\rm BHL}$ is satisfied for sBHs co-rotating with the AGN disk at $r\lesssim$ a few pc, by considering the effect of dust opacity on absorption
\citep[e.g.][]{Toyouchi19} in solar or super-solar metallicity in AGN disks \citep[e.g.][]{Xu18}. 
Also, due to the shadow by the thick accretion disk, photons escape in the vertical direction \citep{Sugimura17}, and then, the HII regions are confined within $r\sim z\lesssim H_{\rm AGN}$. 
Under these conditions, the gas dynamics during capture by the sBH at $r\sim r_{\rm Hill}$ is less affected by the pressure of HII regions. 
However, since the geometry of the system is complex for gas inflow onto sBHs in AGN disks, the effect of radiation may be significantly different. 
We investigate the influence of the possible reduction of the gas density above the sBH by radiation or wind on the regulation of gas accretion in Appendix~$\ref{sec:parameter_dep}$. 
Also, in later episodes, gas inflows during quiescent phases. Due to the quick recombination in the dense AGN disks, we simply assume that gas quickly returns to the neutral state, justifying the resupply timescale (Eq.~\ref{eq:t_re}) assuming 
that gas moves to fill the cavity at the 
local sound speed of the AGN disk $c_{\rm s,AGN}$.

(iii) 
In AGN disks 
outside the dust sublimation radius, 
\begin{eqnarray}
R_{\rm sub}\sim 0.04~{\rm pc}~
\left(\frac{L_{\rm UV}}{10^{43}\,{\rm erg/s}}\right)^{1/2}
\left(\frac{T_{\rm dust}}{1500\,{\rm K}}\right)^{-2.8},
\end{eqnarray}
\citealt{Barvainis1987}, where $L_{\rm UV}$ is the ultraviolet luminosity of the AGN, and $T_{\rm dust}$ is the temperature at which grains are destroyed, the radiation pressure on dust by infrared photons may regulate accretion for high-metallicity environments \citep{Toyouchi19} without preventing gas dynamical friction \citep{Toyouchi20}. 
On the other hand, when the ultraviolet radiation from the sBH is anisotropic due to the shadow by the thick CsBD, this process does not regulate gas accretion \citep{Toyouchi2021}. 
Additionally, while being captured by the sBH, the gas experiences shocks \citep{Lubow1999,Tanigawa2012}, and dust in the CsBD is likely sublimated.

\begin{figure}
\begin{center}
\includegraphics[width=90mm]{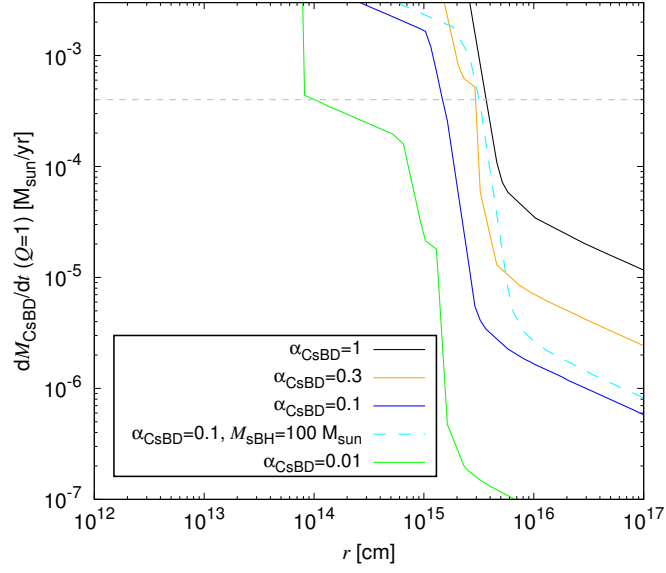}
\caption{
The accretion rate at which the Toomre parameter becomes 1 for several $\alpha_{\rm CsBD}$ and $M_{\rm sBH}$. 
The dashed gray line shows the gas capture rate by an sBH in the fiducial model. 
}
\label{fig:stable_r_md}
\end{center}
\end{figure}

(iv) The captured gas can fragment 
if the CsBD becomes gravitationally unstable. 
Fig.~\ref{fig:stable_r_md} shows the critical accretion rate above which the CsBD becomes Toomre unstable (the Toomre parameter becomes $Q<1$) as a function of $R$. 
Hydrodynamical simulations for proto-planetary disks \citep{Ayliffe2009,Tanigawa2012} 
suggest $r_{\rm CsBD,out}\sim f_{\rm circ} r_{\rm Hill}$
with $f_{\rm circ}\sim 0.1$--$0.4$, corresponding to 
\begin{eqnarray}
r_{\rm CsBD,out}\sim 4\times 10^{15}\,{\rm cm}
\left(\frac{f_{\rm circ}}{0.1}\right) 
\left(\frac{R_{\rm sBH} }{\rm pc}\right) 
\left(\frac{M_{\rm sBH}}{10\,\Msun}\right) ^{1/3} 
\left(\frac{M_{\rm SMBH}}{10^6\,\Msun}\right) ^{-1/3}.
\end{eqnarray}
From Fig.~\ref{fig:stable_r_md}, the CsBD at $r_{\rm CsBD,out}$ becomes unstable roughly at 
\begin{eqnarray}
R_{\rm sBH}\gtrsim {\rm pc}\,
\left(\frac{\alpha_{\rm CsBD}}{0.1}\right)^{c_0}
\left(\frac{f_{\rm circ}}{0.1}\right) ^{-1} 
\left(\frac{M_{\rm sBH}}{10\,\Msun}\right)^{-1/3} 
\left(\frac{M_{\rm SMBH}}{10^6\,\Msun}\right)^{1/3}
\end{eqnarray}
with $c_0\sim 1/2$. 
When the CsBD becomes unstable, $\alpha_{\rm CsBD}$ is expected to be enhanced to $\sim O(1)$ due to gravitational instability, and fragmentation may also occur \citep[e.g.][]{Kratter2016}. 
If $r_{\rm CsBD,out}\gtrsim 10^{16}\,{\rm cm}$, 
the CsBD is unstable even if $\alpha_{\rm CsBD}\sim 1$, and the inflow rate may be reduced to several times $10^{-5}\,\Msun/{\rm yr}$ (black line), although the degree of fragmentation and migration of fragments is not obvious. 
Thus, dynamical instability is expected and may reduce the gas accretion rate only in the outer regions at $R_{\rm sBH}\gtrsim {\rm pc}$.

(v) As described in $\S\,\ref{sec:migration}$, 
a gap is predicted to form around the sBH in an AGN disk 
when $K$ (Eq.~\ref{eq:K_gap}) becomes larger than $\sim 20$. 
This condition is satisfied for $R_{\rm sBH}\lesssim 0.01\,{\rm pc}$ in the fiducial model. 
After the formation of the gap, the growth rate of the sBH is significantly reduced (Figs.~\ref{fig:r_m_noreg} and \ref{fig:r_m}). 
Note that by considering the evolution of many sBHs, deeper gaps are presumed to form due to the sum of their torques at the same radial position, and then the growth of sBHs may be further suppressed especially in $R_{\rm sBH}\lesssim 0.01\,{\rm pc}$. 
Such $N$-body effects on the gap formation will be worth investigating in a future study.

Overall, 
the significant fraction of mass loss via winds \citep[e.g.][]{Jiang+2014,Sadowski+2015}
might reduce the accretion rate onto an sBH, although it is suggested to be modest \citep{Kitaki2021}. 
Also, the inflow rate of the CsBD may be reduced by 
Toomre instability in outer regions of $R_{\rm sBH}\gtrsim {\rm pc}$, and by gap formation in the inner regions with $R_{\rm sBH}\lesssim 0.01\,{\rm pc}$. 
Hence, without additional regulation processes,
sBHs embedded in AGN disks presumably evolve to IMBHs especially for $R_{\rm sBH}\sim 0.01$--$1\,{\rm pc}$ ($\S\,\ref{sec:rapid_growth}$).

\subsection{Influence of winds on cocoon evolution}

\label{sec:inf_wind}

The cocoon dynamics may be also influenced by the winds launched in super-Eddington regimes \citep[e.g.][]{Jiao2015,Kitaki2021}. 
To evaluate this influence, we compare the pressure of the wind to that of the cocoon. 
The ram pressure of the wind at elevation $z$ is 
\begin{eqnarray}
\label{eq:pram_wind}
\rho_{\rm w} v_{\rm w}^2 =\frac{{\dot M}_{\rm w}v_{\rm w}}{\Omega_{\rm w} z^2}
\sim 20\,{\rm erg\,cm^{-3}}~
\left(\frac{{\dot M}_{\rm w}}{3\times 10^{-5}\,{\Msun/{\rm yr}}}\right)
\left(\frac{\beta_{\rm w}}{0.03}\right)
\left(\frac{z}{10^{14}\,{\rm cm}}\right)^{-2}
\left(\frac{\Omega_{\rm w}}{2 \pi}\right)^{-1}
\end{eqnarray}
where $\rho_{\rm w}$ is the wind density, 
$v_{\rm w}=\beta_{\rm w}c$ is the wind velocity, 
and 
$\Omega_{\rm w}$ is the solid angle within which the wind is emitted at the inner region of the accretion disk. 
We assume that ${\dot M}_{\rm w}\sim {\dot M}_{\rm in,vis}-{\dot M}_{\rm sBH}$. 
We approximate the wind velocity to be $\sim ({m}_{\rm sBH}G/r_{\rm trap})^{1/2}\sim  [{\dot M}_{\rm sBH}\eta_{\rm rad}/{\dot M}_{\rm Edd}(M_{\rm sBH})]^{-1/2}c$ \citep[e.g.][]{Kitaki2021}.

On the other hand, the cocoon pressure for ${\tilde L}<1$ is 
\begin{eqnarray}
\label{eq:p_cocoon}
P_c={\tilde L}\theta_0^2 \rho_{\rm AGN} c^2 
\sim 40\,{\rm erg/cm^3}~
\left(\frac{\tilde L}{0.03}\right)
\left(\frac{\theta_{0}}{0.2}\right)^2
\left(\frac{\rho_{\rm AGN}}{4\times 10^{-17}\,{\rm g/cm^3}}\right).
\end{eqnarray}
By equating Eqs.~\eqref{eq:pram_wind} and \eqref{eq:p_cocoon}, 
the pressures become comparable at the equilibrium height of  
\begin{eqnarray}
\label{eq:zp_wind_cocoon}
z_{\rm eq} \sim 8\times 10^{13}\,{\rm cm}~
\left(\frac{{\dot M}_{\rm w}}{3\times 10^{-5}\,{\Msun/{\rm yr}}}\right)^{1/2}
\left(\frac{\beta_{\rm w}}{0.03}\right)^{1/2}
\left(\frac{\Omega_{\rm w}}{2\pi}\right)^{-1/2}
\left(\frac{{\tilde L}}{0.03}\right)^{-1/2}
\left(\frac{\theta_{0}}{0.2}\right)^{-1}
\left(\frac{\rho_{\rm AGN}}{4\times 10^{-17}\,{\rm g/cm^3}}\right)^{-1/2}\,.
\end{eqnarray}
In the fiducial setting, 
$z_{\rm eq}$ is somewhat smaller than the truncation radius (Eq.~\ref{eq:r_tru}), i.e. by a factor of several in no gap-forming regions (cyan line in Fig.~\ref{fig:r_vari3}~a). In this case ($z_{\rm eq}<r_{\rm tru}$), the cocoon can interact with and heat the CsBD at $r=r_{\rm tru}$ without being inhibited by the wind pressure. 

Even when $z_{\rm eq}>r_{\rm tru}$, 
the truncation radius is presumably unaffected by the wind. The wind pressure decreases with $r$, and the cocoon may be able to reach the disk midplane at $r\gtrsim z_{\rm eq}>r_{\rm tru}$. Then, the shock propagates to the inner part ($r_{\rm tru}<r<z_{\rm eq}$) of the CsBD, and eventually arrives at $r=r_{\rm tru}$. Also, 
a large fraction of the momentum by the wind is presumably confined within some solid angle from a polar axis \citep{Jiao2015,Kitaki2021}, 
and then, winds do not affect cocoon dynamics around the midplane.

When the sBH is slowly spinning, the jet efficiency ($\eta_{\rm j}$) becomes low and feedback is presumed to be dominated by the wind at some point. 
If $\eta_{\rm j}\sim a_{\rm sBH}^2$ is lower than $\sim f_{\rm w}(v_{\rm w}/c)^2 \sim f_{\rm w}({\dot m}_{\rm sBH}\eta_{\rm rad})^{-1}$, where $f_{\rm w}={\dot M}_{\rm w}/{\dot M}_{\rm cap}$ is the fraction of the wind loss rate over the capture rate, the luminosity of the jet roughly becomes lower than that of the wind. 
From this relation, we presume that regulation of accretion is mainly conducted by winds for
$a_{\rm sBH}\lesssim [f_{\rm w}/({\dot m}_{\rm sBH}\eta_{\rm rad})]^{1/2}$, although understanding of this transition in detail would require performing hydrodynamical simulations. 
We conclude that the effect of the wind on the cocoon evolution can be neglected for a rapidly accreting and spinning sBH.

\begin{figure}
\begin{center}
\includegraphics[width=150mm]{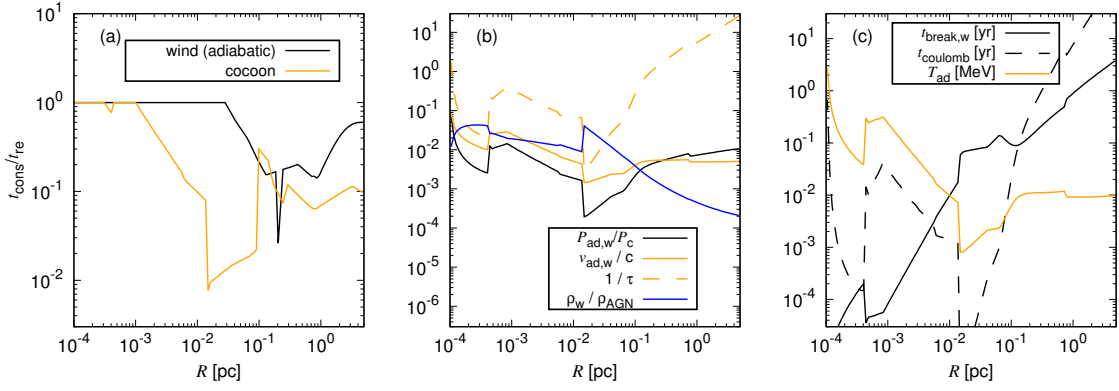}
\caption{
Various properties related to the shock produced by the wind as a function of $R$. 
{\it Left} panel~(a): 
The reduction rate of accretion onto the sBH by the cocoon (orange) and by the wind shock 
without cooling (black). 
{\it Middle} panel~(b): 
The ratio of the thermal pressure with the temperature $T_{\rm ad}$ of the wind shock over the internal pressure of the cocoon (black), 
the dimensionless shock velocity at breakout assuming adiabatic evolution (solid orange, $v_{\rm ad,w}/c$), 
the dimensionless photon diffusion velocity from the AGN disk (dashed orange, $1/\tau=1/2\rho_{\rm AGN}H_{\rm AGN}\kappa_{\rm AGN}$), and 
the ratio of the wind density at $z=H_{\rm AGN}$ to the AGN density (blue). 
{\it Right} panel~(c): 
The breakout timescale of the wind assuming adiabatic evolution (solid black), 
the Coulomb loss timescale for protons by collisions with electrons in the shock (dashed black),  
and the proton temperature of the shocked fluid assuming inefficient cooling 
(orange). 
In the case of the wind shock, accretion is regulated at $R\gtrsim 0.1~{\rm pc}$, where cooling is inefficient. 
}
\label{fig:wind}
\end{center}
\end{figure}

\subsection{Regulation of accretion by winds/outflows}

\label{sec:wind_feedback}

In this section, we evaluate regulation of accretion onto the sBH by radiation-driven winds (outflows) launched from a rapidly accreting CsBD. 
Fig.~\ref{fig:wind} shows various properties of the shock produced by the wind. 
The wind density without interactions is given by $\rho_{\rm w} =\frac{{\dot M}_{\rm w}}{\Omega_{\rm w} \beta_{\rm w} c z^2}$, 
which is lower than the disk density at the scale height of the AGN disk (blue line in Fig.~\ref{fig:wind}~b). 
This suggests that 
the wind is decelerated before breakout from the AGN disk. 
When radiation is inefficient, 
the internal pressure of the shocked fluid is dominated by 
the thermal energy ($P_{\rm ad,w}=k_{\rm B} T_{\rm ad}n_{\rm AGN}$, black in panel~b), 
where $T_{\rm ad}=0.7 L_{\rm w} t_{\rm break,w}/(3/2 k_{\rm B}n_{\rm AGN}V_{\rm sh,w})$ is the proton temperature at breakout assuming negligible radiative cooling 
(e.g., \citealt{McKee1977}, orange line in panel~c), 
$L_{\rm w}={\dot M}_{\rm w}v_{\rm w}^2$ is the wind luminosity, 
$t_{\rm break,w}=H_{\rm AGN}/v_{\rm ad,w}$ is the breakout timescale of the wind shock from the AGN disk assuming adiabatic evolution (solid black in panel~c), 
$v_{\rm ad,w}$ is the shock velocity at breakout assuming adiabatic evolution, 
and $V_{\rm sh,w}$ is the volume of the shocked region at the breakout.

When the photon diffusion velocity from the AGN disk ($c/\tau$) exceeds 
the velocity of the wind shock ($v_{\rm sh,w}$), 
photons can escape from each shocked shell, 
where $\tau=2\rho_{\rm AGN}H_{\rm AGN}\kappa_{\rm AGN}$ is the optical depth of the AGN disk.  
Since $v_{\rm ad,w}$ is the upper limit for $v_{\rm sh,w}$, 
the diffusion of photons is faster than the shock velocity ($v_{\rm sh,w}\leq v_{\rm ad,w}<c/\tau$, solid and dashed orange lines in Fig.~\ref{fig:wind}~b). 
Then, 
the energy loss rate by cooling may reduce the internal pressure of the shocked regions. 
The cooling of protons becomes efficient when 
$t_{\rm break,w}$ is shorter than the Coulomb loss timescale ($t_{\rm coulomb}=m_p m_e c_{\rm s,ad}^3 3^{3/2}/(8\pi e^4 n_{\rm AGN}{\rm log}\Lambda)$, e.g., \citealt{Dermer2009}, dashed black in panel~c), where $m_p$ and $m_e$ are the proton and electron masses, ${\rm log}\Lambda={\rm log}(r_{\rm d}/r_{\rm c})$ is the Coulomb logarithm, $r_{\rm d}$ is the Debye length, $r_{\rm c}=e^2/k_{\rm B} T_{\rm ad}$ is the collision parameter, $e$ is the elementary charge, and $c_{\rm s,ad}$ is the sound speed of protons with the temperature $T_{\rm ad}$. 
In the fiducial model, $t_{\rm break,w}$ exceeds $t_{\rm coulomb}$ at $0.1~{\rm pc}\gtrsim R \gtrsim 0.003~{\rm pc}$ and $R \sim 3\times 10^{-4}~{\rm pc}$. 
In these regions, cooling of protons and loss of the shock energy (and reduction of the internal pressure) are expected to become efficient, 
since the Coulomb interaction converts the thermal energy of protons to that of electrons, 
free-free emission converts the thermal energy of electrons to the radiation energy, 
with Compton scattering  contributing at $R\lesssim 10^{-2}$--$10^{-3}~{\rm pc}$, where the electron temperature 
and the Compton parameter \citep[e.g.][]{Rybicki1979} are presumed to be high. 
However, detailed analyses of the cooling processes require to model the evolution of electrons considering interactions of a collisionless plasma, which are complicated and uncertain, and also cooling has a minor influence on the reduction of accretion as discussed below. For simplicity, we leave a more detailed analysis of cooling to a future study.

If the shocked fluid is adiabatic, 
the internal energy of the shock produced by the wind is not negligible compared to the cocoon's internal energy (black line in panel~b). This is because even though the luminosity of the wind is lower than that of the jet by a factor of $\sim c^2/v_{\rm w}^2$ assuming a same conversion efficiency ($\eta_{\rm j}\sim {\dot M}_{\rm w}/{\dot M}_{\rm sBH}$), there is a long time for the wind to deposit energy.
$P_{\rm ad,w}$ is lower than $P_{\rm c}$ mostly because the volume of the wind shock is larger than that of the cocoon by $\sim \theta_{0}^{-2}$. 
Note that the dependence of $t_{\rm cons}$ and $t_{\rm re}$ on $\theta_0$ is roughly the same, so $\theta_0$ does not contribute to the reduction rate. 
By performing a similar analysis to the 
JFM 
using $P_{\rm ad,w}$ (solid black in panel~b), we can estimate the reduction rate of accretion onto the sBH by the wind assuming adiabatic evolution of the wind shock
(solid black line in panel~(a) of Fig.~\ref{fig:wind}). 
By comparing this reduction to that by the cocoon (orange line), 
we can see that the reduction rate of accretion onto the sBH by the wind is a minor contribution inside $R\lesssim 10^{-1}~{\rm pc}$, 
while it can become comparable to that by the cocoon at $R\gtrsim 10^{-1}~{\rm pc}$ if cooling is inefficient. 
As discussed above, the internal pressure of the wind shock can be reduced by cooling within $R\lesssim 0.1~{\rm pc}$, while that at $R\gtrsim 10^{-1}~{\rm pc}$ is less influenced. 
Thus, the reduction rate of accretion by the wind is roughly given by the black line in Fig.~\ref{fig:wind}(a). 
When the sBH is 
rapidly spinning 
and produces a strong BZ jet, the jet and the cocoon propagate faster and the 
JFM 
dominates as discussed in the previous section. 
On the other hand, for an sBH that is not rapidly spinning, 
the wind can play a dominant role in regulating accretion at $R\gtrsim 10^{-1}\,{\rm pc}$. 
If accretion of sBHs is mainly regulated by the wind shock, 
the critical radius for the ejection of captured gas is larger by a factor of $\sim P_{\rm c}/P_{\rm ad,w}$ (black line in Fig~\ref{fig:wind}~b) 
compared to that by the cocoon, due to less efficient heating by the winds. 
Then, the depletion of gas is not expected even for less massive SMBHs of $M_{\rm SMBH}\sim 10^4$--$10^5~\Msun$, which is different from the JFM 
($\S\,\ref{sec:discussion}$). 
Thus, quasar observations for less massive SMBHs with high accretion rates \citep{Greene07} can be used to distinguish the feedback processes for accretion onto sBHs in this environments.

\subsection{Influence of the cocoon on migration and gas dynamical friction}

We next discuss the influence of the 
JFM 
on migration \citep[e.g.][]{Ostriker99}. 
The migration of sBHs in an AGN disk is caused by resonant Lindblad and corotation torques \citep[e.g.][]{Armitage2007_note}, 
which are mainly contributed by gas outside and inside the Hill radius, respectively. 
As $r_{\rm dep} < H_{\rm AGN} \lesssim r_{\rm Hill}$, the corotation torque is predicted to be significantly affected, while the Lindblad torque is not. 
To understand how the corotational torque is modified, 
hydrodynamical simulations need to be performed.

We also investigate whether the depth of the gap is influenced by the cocoon evolution. 
Since the cocoon can eject gas which is captured by the sBH, 
the maximum ejection rate is this capture rate. 
We confirmed that the gas capture does not influence the gap-opening timescale and the gap depth in the fiducial cases 
since the gravitational torque from sBHs more efficiently opens the gap compared to the gas capture. 
The gravitational torque from an sBH is $T_{\rm sBH} \sim (M_{\rm sBH}/M_{\rm SMBH})^2 (R_{\rm sBH}/H_{\rm AGN})^3 R_{\rm sBH}^4\Omega_{\rm Kep}^2 \Sigma_{\rm AGN}$ \citep{Kanagawa18}, 
where $\Sigma_{\rm AGN}$ is the surface density of the AGN disk, 
and the angular momentum of gas within the annular gap is $\Delta J \sim 2\pi R_{\rm sBH}^2 \Delta_{\rm gap} \Omega_{\rm Kep} \Sigma_{\rm AGN}$, where $\Delta_{\rm gap}$ is the gap width. 
By comparing the timescale of gap opening by the gravitational torque ($t_{\rm torque}\sim \Delta J/ T_{\rm sBH}$) 
to that by gas capture ($t_{\rm cap}\sim M_{\rm gap}/{\dot M}_{\rm cap}$), where $M_{\rm gap}\sim 2\pi R_{\rm sBH} \Delta_{\rm gap}\Sigma_{\rm AGN}$ is the disk mass within a gap, we can see that
the gravitational torque dominates when $(M_{\rm sBH}/M_{\rm SMBH})^2\gtrsim (f_{\rm c}/2)(H_{\rm AGN}/R_{\rm sBH})^3 (\Delta_{\rm gap}/R_{\rm sBH})(R_{\rm w}R_{\rm h}/R_{\rm sBH}^2)$. 
By roughly assuming $\Delta_{\rm gap}\sim R_{\rm Hill}\sim H_{\rm AGN}$, 
the condition can be simplified to $M_{\rm sBH}/M_{\rm SMBH}\gtrsim (H_{\rm AGN}/R_{\rm sBH})^3$, which is usually satisfied unless $M_{\rm SMBH}$ and $H_{\rm AGN}/R_{\rm sBH}$ are high. 
In the cases that the AGN disk is thick, a gap can be efficiently created due to gas capture by sBHs instead of gravitational torques.

Similarly, we consider the influence of the 
JFM 
on gas dynamical friction \citep[e.g.][]{Ostriker99}. Since gas dynamical friction arises from torques from gas around the Bondi-Hoyle-Lyttleton radius, 
and $r_{\rm dep}<H_{\rm AGN} \ll r_{\rm BHL}$ in the AGN disk, 
we conclude that gas dynamical friction is not significantly affected by the cocoon evolution, which ejects gas within $r_{\rm dep}$.

Hence, the prescription for migration of sBHs (Eq.~\ref{eq:typeI_II})
and the number of sBHs captured by an AGN disk (Fig.~\ref{fig:dep_m}) ignoring the 
JFM 
are 
presumably justified.

\begin{figure}
\begin{center}
\includegraphics[width=110mm]{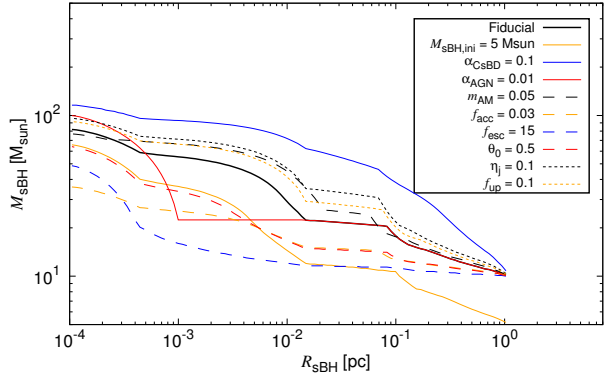}
\caption{
Same as in the lower panel of Fig.~\ref{fig:r_m}, 
but with parameters varied from the fiducial model as indicated in the legend. 
}
\label{fig:m_dep}
\end{center}
\end{figure}

\begin{figure}
\begin{center}
\includegraphics[width=110mm]{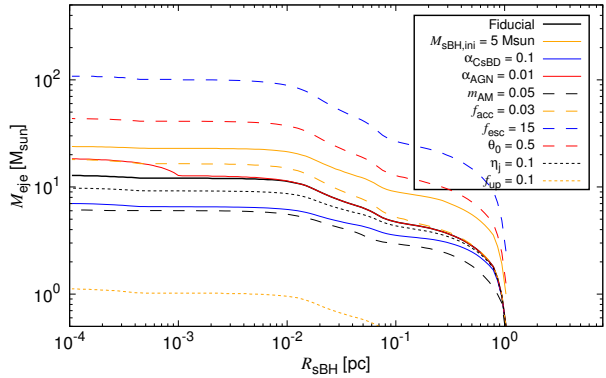}
\caption{
Same as Fig.~\ref{fig:m_dep}, 
but for the total ejected mass from the AGN disk 
due to the 
JFM 
by the sBH migrating to $R_{\rm sBH}$. 
}
\label{fig:meje_dep}
\end{center}
\end{figure}

\section{Parameter dependence}

\label{sec:parameter_dep}

Finally, we discuss the parameter dependence of the regulation of gas accretion. 

\subsection{Final mass}

\label{sec:parameter_dep_final_mass}

We study how the evolution of the sBH mass depends on the model parameters. 
Fig.~\ref{fig:m_dep} shows $M_{\rm sBH}$ (Eq.~\ref{eq:macc_tagn}) for various choices of parameters. 
We can see that $M_{\rm sBH}(R_{\rm sBH})$ is very parameter-dependent.
As $M_{\rm sBH,ini}$ or $f_{\rm acc}$ decrease, the average accretion rate onto the sBH is reduced (solid and dashed orange lines, Eq.~\ref{eq:md_bhl_edd_ave}), which lowers the final mass of the sBH. 
For low $\alpha_{\rm CsBD}$ (solid blue), the surface density of the CsBD is high, in which case the truncation radius is large, and the viscous timescale ($\sim t_{\rm cons}$) at a fixed radius is long, both of which enhance the duty cycle ($t_{\rm cons}/t_{\rm re}$) and the growth rate of the sBH. 
A high value for $f_{\rm ext}$ or $\theta_0$ increase $r_{\rm dep}$, which reduces the duty cycle and the growth rate (dashed blue and red). 
When the angular momentum transfer parameter for the AGN disk is lower (solid red and dashed black), the surface density of the disk is higher, 
which enhances the accretion rate and the mass of the sBHs at a fixed location $R_{\rm sBH}$. 
When the efficiency for producing the jet is weak (dotted black), 
the accretion becomes efficient as the ejected mass of the CsBD is reduced. 
We also investigate the case in which the gas density above the sBH is reduced to $\rho_{\rm AGN} f_{\rm up}$, possibly due to radiation and mechanical feedback. In this case, the pressure of the cocoon is reduced, and the regulation of accretion is weakened (dotted orange). 
Hence, 
the regulation of gas accretion by the cocoon strongly depends on many parameters (Eq.~\ref{eq:md_bhl_edd_ave}). 

Fig.~\ref{fig:meje_dep} shows the total ejected mass from the AGN disk due to the 
JFM 
by the sBH migrating to $R_{\rm sBH}$. We assume the ejected mass per one cocoon evolution to be $M_{\rm eje}=\pi r_{\rm dep}^2 H_{\rm AGN} \rho_{\rm AGN}$. 
From Fig.~\ref{fig:meje_dep}, the ejected mass from the AGN disk is smaller than the accreted mass onto the sBH (Fig.~\ref{fig:dep_m}) except for the large $f_{\rm esc}$ and $\theta_0$ cases (dashed blue and red lines), when gas in larger regions is ejected by the feedback. 
Even in such cases of efficient ejection, the ejected mass from the AGN disk does not contribute to the depletion of gas inflow.

\subsection{Inclined cases}

In our modeling as described in the main text, we assume that the jet direction is the same as the angular momentum direction of the AGN disk. 
On the other hand, 
when the spin direction of the sBH is misaligned with respect to the angular momentum direction of the AGN disk 
\citep[e.g.][for the inclined JFM in common envelope evolution]{Schreier2019}, 
it takes a longer time for the jet-cocoon to break out. 
Then, the depletion region by the cocoon is effectively enhanced to $r_{\rm dep}=f_{\rm ext}H_{\rm AGN}\theta_{\rm c}/{\rm cos}i$, where $i$ is the inclination angle of the jet with respect to the orbital angular momentum of the AGN disk.

The sBH spins are often randomly oriented if enhanced by mergers in gap-forming regions \citep{Tagawa20b_spin,Tagawa2021_hierarchical}. In this case, the resupply timescale increases by a factor of a few on average, which effectively reduces the average accretion rate by a similar factor. 
On the other hand, if sBH spins evolve due to mergers in migration traps \citep{Bellovary16,Yang19b_PRL,Yang20_gap} or gas accretion \citep[e.g.][]{Safarzadeh20_GW190521} in the AGN disk, $i$ is presumably close to zero, and the formulae for the depletion radius (Eq.~\ref{eq:r_dep}) and the resupply timescale (Eq.~\ref{eq:t_re}) do not need revisions.

When the inclination exceeds $\pi/4$, the jet likely collides with thick regions of the CsBD within the trapping radius. In this case, geometrically-thick inner regions are thermalized as a cocoon, and ejected from the sBH. This cocoon possibly escapes without ejecting thin regions of the CsBD since shocked materials tend to proceed towards low pressure regions \citep{Kompaneets1960}. The overall picture of sBH growth should not be significantly modified in such cases.  Since the probability is low (of about the aspect ratio of the disk), 
we ignore the situations in which the jet collides with geometrically thin regions of the CsBD, in which case most of the CsBD is probably ejected. 

Meanwhile, if sBHs are formed in-situ \citep{Fahrion2021}, their formation may be accompanied by long gamma-ray bursts \citep{Zhu2021_Cocoon_NSMs,Perna2021_GRBs,Jermyn2021}, supernova explosions \citep{Grishin2021,Zhu2021_Neutrino}, or energetic transients associated with the accretion-induced collapse of neutron stars and white dwarfs \citep{McKernan2020,Perna2021_AICs,Zhu2021_WD_AIC}. Thus, the first episode for gas ejection may be distinct from the later episodes. 
The observational signatures related to a jet and a cocoon will be investigated in a forthcoming paper.

\section{Variables}

\subsection{Parameter set}

The fiducial model parameters are listed in Table~\ref{table:parameter_fiducial}.

\begin{table*}
\begin{center}
\caption{Fiducial values of our model parameters.}
\label{table:parameter_fiducial}
\hspace{-5mm}
\begin{tabular}{c|c}
\hline 
Parameter & Fiducial value \\
\hline\hline
Radial distance of the sBH from the SMBH & $R_{\rm sBH}=1\,{\rm pc}$\\\hline
Jet energy conversion efficiency & $\eta_{\rm j}=0.5$\\\hline
Initial mass of the sBH & $M_{\rm sBH,ini}=10\,{\Msun}$\\\hline
Mass of the SMBH & $M_{\rm SMBH}=10^6\,{\Msun}$\\\hline
Gas inflow rate from the outer boundary of the AGN disk in units of the Eddington rate for $M_{\rm SMBH}$& ${\dot m}_{\rm SMBH}=0.1$\\\hline
Outer boundary of the AGN disk & $R_{\rm out}=5\,{\rm pc}$\\\hline
Angular momentum transfer parameter in the outer regions of the AGN disk & $m_{\rm AM}=0.15$\\\hline
Viscous parameter of the AGN disk & $\alpha_{\rm AGN}=0.1$\\\hline
Ratio of the sBH accretion rate to the gas capture rate & $f_{\rm acc}=0.1$\\\hline
Opening angle at the base of the jet & $\theta_{\rm 0}=0.2$\\\hline
Radiative efficiency & $\eta_{\rm rad}=0.1$\\\hline
Viscous parameter of the CsBD & $\alpha_{\rm CsBD}=0.3$\\\hline
Ratio of the outer radius of the CsBD to the Hill radius
& $f_{\rm circ}=0.1$ \\\hline

\end{tabular}
\end{center}
\end{table*}

\subsection{Notation}

The notations of variables are listed in Tables~\ref{table_notation} and \ref{table_notation2}. 

\begin{table*}
	\caption{
		Notation. 
        	}
\label{table_notation}
\hspace{-0.0mm}
\begin{tabular}{p{2.5cm}|p{5cm}||p{2.5cm}|p{5cm}}
\hline
Symbol&Description&Symbol&Description\\\hline
$r$, $z$, $R$, $Z$& 
The cylindrical coordinates $(z,r)$ and $(Z,R)$ with $z=r=0$ and $Z=R=0$ to be the positions of the sBH and the SMBH, respectively 
& 
$R_{\rm sBH}$& The radial position of the sBH from the SMBH
\\\hline

$r_{\rm BHL}$, $r_{\rm Hill}$& 
The Bondi-Hoyle-Lyttleton radius, the Hill radius
&
$r_{\rm w}={\rm min}(r_{\rm BHL},\,r_{\rm Hill})$, $r_{\rm h}={\rm min}(r_{\rm w},\,H_{\rm AGN})$
& The capture radius and height of gas by an sBH 
\\\hline

$r_{\rm tru}$, $r_{\rm dep}$, $r_{\rm c}$& The truncation radius outside which the CsBD is truncated, and the depletion radius within which the AGN gas is ejected, the $r$-direction extent of the cocoon at the jet breakout & 
$r_{\rm CsBD,out}$, $r_{\rm vis}$& 
The outer radius of the CsBD, the radial distance from the sBH at which the viscous timescale becomes equal to the resupply timescale ($t_{\rm re}$)
\\\hline

$H_{\rm AGN}$& The scale height of the AGN disk&
$z_{\rm eq}$ & The equilibrium height at which the pressures by the cocoon and the wind becomes equal
\\\hline

$v_\mathrm{CsBD,Kep}(r)$ & The Keplerian velocity of the CsBD at the distance $r$ from the sBH &
$\Omega_{\rm Kep}$& The Keplerian angular velocity around the SMBH
\\\hline

$c_{\rm s,AGN}$, $v_{\rm sBH}$, $v_{\rm sh}$ & The sound velocity of the AGN disk, the velocity of the sBH with respect to the local AGN motion, and the shear velocity at the capture radius ($r_{\rm w}$)
&
$v_{\rm w}=\beta_{\rm w}c$, ${\dot M}_{\rm w}$, $\Omega_{\rm w}$& The velocity, outflow rate, and solid angle of the wind 
\\\hline

$M_{\rm sBH}$, $M_{\rm SMBH}$& The mass of the sBH and the SMBH &
$M_{\rm sBH,ini}$, $M_{\rm sBH,fin}$ & 
The mass of the sBH at $t=0$ and $t=t_{\rm AGN}$
\\\hline

$M_{\rm annu}$, $M_{\rm ej}$& The gas mass within annulus with the width of $r_{\rm Hill}$ around the sBH, the ejection mass by the cocoon evolution & 
$M_{\rm CsBD,tru}$ & The CsBD mass after the truncation (with the radius $r_{\rm tru}$) by the evolution of the cocoon
\\\hline

${\dot M}_{\rm cap}$, ${\dot M}_{\rm in,vis}$, ${\dot M}_{\rm sBH}$& 
The capture rate (Eq.~\ref{eq:md_bondi}), 
the inflow rate within $r_{\rm vis}$,
the accretion rate onto an sBH 
&
${\dot M}_{\rm Edd}(M)$&
The Eddington accretion rate onto a BH with the mass $M$ 
\\\hline

${\dot M}_{\rm sBH,tot}$& 
The depletion rate by all sBHs embedded in an AGN disk
&
${\dot M}_{\rm CsBD,Kep}(<r)$
&
The cumulative mass accretion rate of accreting gas 
after redistributing the radial distances $r$ where their specific angular momentum matches that of the local Keplerian rotation
\\\hline

${{\dot m}}_{\rm SMBH}={\dot M}_{\rm SMBH,in}/{\dot M}_{\rm Edd}$ 
& The gas inflow rate from the outer boundary of the AGN disk in units of the Eddington rate for $M_{\rm SMBH}$& 
${\dot M}_{\rm SMBH,in}$ & The inflow rate of the AGN disk from the outer boundary ($R_{\rm out}$)
\\\hline

$\Sigma_{\rm CsBD}$ & The surface density of the CsBD&
$\Sigma_{\rm disk}, \Sigma_{\rm disk,min}$ & The surface density of the AGN disk and that after a gap forms\\\hline

$\rho_{\rm AGN}$, $\Sigma_{\rm AGN}$& The density and surface density of the AGN disk &
$L_{\rm Edd}(M)$, 
$\eta_{\rm rad}$& The Eddington luminosity for the mass $M$, the conversion efficiency to radiation 
\\\hline

$f_{\rm acc}={\dot M}_{\rm sBH}/{\dot M}_{\rm cap}$& 
The reduction factor of the captured gas&
$f_{\rm acc,eff}={\dot M}_{\rm sBH,ave}/{\dot M}_{\rm cap}$& The averaged reduction factor of the captured gas
\\\hline

$f_{\rm circ}$ & The ratio of the outer radius of the CsBD over the Hill radius
&
$f_{\rm ext}=r_{\rm dep}/r_{\rm c}$& The fraction that the cocoon proceeds to the $r$-direction after the breakout  
\\\hline

$f_{\rm w}={\dot M}_{\rm w}/{\dot M}_{\rm cap}$ & 
The fraction of the wind loss rate over the capture rate&
$N_{\rm AGN,sBH}$& 
The typical number of sBHs embedded in an AGN disk
\\\hline

$t_{\rm re}$, $t_{\rm cons}$ & The resupply timescale of ejected gas ejected by the cocoon evolution, the consumption timescale of the bound CsBD after the cocoon evolution&
$t_{\rm I}$, $t_{\rm I,II}$& The type I and type I/II migration timescale
\\\hline

$t_{\rm break}$& The break out timescale of the cocoon & 
$t_{\rm gap,ej}$ & The timescale of the gas ejection of the mass, $M_{\rm annu}$ \\\hline

$t_{\rm AGN}$& 
The typical total AGN phases in a galaxy 
& 
$t_{\rm mig}$ & The migration timescale in disks\\\hline

${\tilde L}$ & The ratio between the energy density of the jet and the rest-mass energy density of the AGN disk 
&
$L_{\rm j}$, $\eta_{\rm j}$ & The luminosity of the jet, the conversion efficiency to jet
\\\hline

$\beta_{\rm h}$, $\beta_{\rm c}$, $\Gamma_{\rm j}$& The head velocity and the lateral expansion velocity of the cocoon, the Lorentz factor just below the head& 
$E_{\rm c}$, $V_{\rm c}$, $P_{\rm c}$ & The energy, volume, and pressure of the cocoon \\\hline

$\theta_0$, $\theta_{\rm c}$& The opening angle of injected jet and the cocoon& 
$a_{\rm sBH}$& The dimensionless spin of the sBH
\\\hline

$\alpha_{\rm CsBD}$, $\alpha_{\rm AGN}$& The viscous parameter for the CsBD and the AGN disk & 
$m_{\rm AM}$& 
The angular momentum transfer parameter for outer regions of the AGN disk  \\\hline

$\kappa_{\rm CsBD}$, 
$\kappa_{\rm AGN}$& The opacity of the CsBD and the AGN disk 
&
$K$ & The parameter representing the depth of a gap
\\\hline

\end{tabular}
\end{table*}

\begin{table*}
	\caption{
		Notation. 
        	}
\label{table_notation2}
\hspace{-0.0mm}
\begin{tabular}{p{2.5cm}|p{5cm}||p{2.5cm}|p{5cm}}
\hline
Symbol&Description&Symbol&Description\\\hline

$c$ & The speed of light & 
$k_{\rm B}$& The Boltzmann constant
\\\hline

$v_{\rm esc}$& The escape velocity from the SMBH &
$i$& The inclination angle between the jet and the orbital angular momentum of the AGN disk 
\\\hline

$\nu$ & The viscosity& 
$\alpha_{\rm eff}=\nu /(c_{\rm s,AGN} H_\mathrm{AGN})$& The effective $\alpha$ parameter 
\\\hline

$\rho_{\rm w}$ & The wind density& 
$L_{\rm w}$& The wind luminosity
\\\hline

$v_{\rm sh,w}$, $v_{\rm ad,w}$& The shock velocity, and that assuming the adiabatic evolution at breakout from the AGN disk&
$T_{\rm ad}$
& The gas temperature for the shocked fluid at breakout from the AGN disk assuming adiabatic evolution 
\\\hline

$P_{\rm ad}$& The thermal pressure with the temperature $T_{\rm ad}$& 
$V_{\rm sh,w}$& The volume of the shocked region at breakout from the AGN disk
\\\hline

$t_{\rm break,w}$& The breakout timescale from the AGN disk of the wind shock assuming adiabatic evolution &
$t_{\rm coulomb}$& The Coulomb loss timescale \\\hline

$t_{\rm acc}$ & The accretion timescale & 
$M_{\rm crit}$ & The critical mass for the sBH at which $t_{\rm acc}=t_{\rm I}$ 
\\\hline

$m_p$, $m_e$& The proton and electron masses& 
${\rm log}({\Lambda})$& The Coulomb logarithm \\\hline

$r_{\rm d}$, $r_{\rm c}$& The Debaye length and the collision length& 
$e$& The elementary charge \\\hline

$\tau=2\rho_{\rm AGN}H_{\rm AGN}\kappa_{\rm AGN}$& The optical depth of the AGN disk& 
$y_{\rm comp}$& The Compton parameter \\\hline

\end{tabular}
\end{table*}

\clearpage
\newpage

\bibliographystyle{aasjournal}
\bibliography{agn_bhm}

\end{document}